%% file: ms.tex
\documentclass[12pt, draftclsnofoot, onecolumn]{IEEEtran}

\usepackage[utf8]{inputenc} 
\usepackage[T1]{fontenc}
\usepackage{url}
\usepackage{ifthen}
\usepackage{cite}
\usepackage{amssymb}
\usepackage[cmex10]{amsmath} 

\usepackage{graphicx}
\usepackage{xcolor}
\usepackage{epstopdf}

\usepackage{subcaption}

\graphicspath{ {./Figures/} }

\newcommand{\vX}{\mathbf{X}}
\newcommand{\vXH}{\mathbf{\hat{X}}}

\newcommand{\vY}{\mathbf{Y}}
\newcommand{\vH}{\mathbf{H}}
\newcommand{\vI}{\mathbf{I}}
\newcommand{\vYQ}{\mathbf{Y}_Q}
\newcommand{\vYV}{\mathbf{Y}_V}
\newcommand{\vYVH}{\mathbf{\bar{Y}}_V}
\newcommand{\vN}{\mathbf{N}}
\newcommand{\bvY}{\bar{Y}}

\newcommand{\vx}{\mathbf{x}}
\newcommand{\vy}{\mathbf{y}}

\newcommand{\vyQ}{\mathbf{y}_Q}
\newcommand{\bvyQ}{\bar{\mathbf{y}}_Q}
\newcommand{\bvx}{\bar{\mathbf{x}}}

\newcommand{\vk}{\mathbf{m}}
\newcommand{\vl}{\mathbf{n}}

\newcommand\inv[1]{#1\raisebox{1.15ex}{$\scriptscriptstyle-\!1$}}
\newcommand{\norm}[1]{\left\lVert#1\right\rVert}

\begin{document}

\title{All-Digital LoS MIMO with Low-Precision Analog-to-Digital Conversion}

\author{Ahmet~Dundar~Sezer,~\IEEEmembership{Member,~IEEE,}
        and~Upamanyu~Madhow,~\IEEEmembership{Fellow,~IEEE}
\thanks{This work was supported in part by ComSenTer, one of six centers in JUMP, a Semiconductor Research Corporation (SRC) program sponsored by DARPA. Use was made of computational facilities purchased with funds from the National Science Foundation (CNS-1725797) and administered by the Center for Scientific Computing (CSC). The CSC is supported by the California NanoSystems Institute and the Materials Research Science and Engineering Center (MRSEC; NSF DMR 1720256) at UC Santa Barbara.}
\thanks{A. D. Sezer and U. Madhow are with the Department
of Electrical and Computer Engineering, University of California Santa Barbara, CA, 93106 USA (e-mail: adsezer@ece.ucsb.edu, madhow@ece.ucsb.edu)} 
}

\maketitle


\input{abstract}

\begin{IEEEkeywords}
Millimeter wave, THz, LoS, MIMO, ADC, quantization, quantizer design, spatial demultiplexing.
\end{IEEEkeywords}

\IEEEpeerreviewmaketitle

\input{intro}

\input{related_work}

\input{system_model_prob_formulation}

\input{2by2System}

\input{4by4System}

\input{detection_virtual_quantization}

\input{conclusion}


\bibliographystyle{IEEEtran}
\bibliography{quantization}

\end{document}

%% file: abstract.tex
\begin{abstract}
Line-of-sight (LoS) multi-input multi-output (MIMO) systems exhibit attractive scaling properties with increase in carrier frequency: for a fixed form factor and range, the spatial degrees of freedom increase quadratically for 2D arrays, in addition to the typically linear increase in available bandwidth. In this paper, we investigate whether modern all-digital baseband signal processing architectures can be devised for such regimes, given the difficulty of analog-to-digital conversion for large bandwidths. We propose low-precision quantizer designs and accompanying spatial demultiplexing algorithms, considering $2 \times 2$ LoS MIMO with QPSK for analytical insight, and $4 \times 4$ MIMO with QPSK and 16QAM for performance evaluation. Unlike prior work, channel state information is utilized only at the receiver (i.e., transmit precoding is not employed). We investigate quantizers with regular structure whose high-SNR mutual information approaches that of an unquantized system. We prove that amplitude-phase quantization is necessary to attain this benchmark; phase-only quantization falls short. We show that quantizers based on maximizing per-antenna output entropy perform better than standard Minimum Mean Squared Quantization Error (MMSQE) quantization. For spatial demultiplexing with severely quantized observations, we introduce the novel concept of virtual quantization which, combined with linear detection, provides reliable demodulation at significantly reduced complexity compared to maximum likelihood detection.
\end{abstract}

%% file: intro.tex
\section{Introduction}\label{sec:intro}

\IEEEPARstart{L}{ine} of sight (LoS) multi-input multi-output (MIMO) communication is well-matched to higher carrier frequencies in the millimeter (mm) wave and THz bands because of the attractive scaling of spatial degrees of freedom (DoF) and bandwidth. For 1D apertures with a horizontal distance, or range, $R$ between transmit and receive arrays having lengths of $L_T$ and $L_R$, the number of spatial DoF based on information-theoretic considerations, given by \cite{Torkildson}
\begin{gather}\label{eq:DoF_1D}
DoF \approx \frac{L_T L_R}{R \lambda } + 1 \,
\end{gather}
scales inversely with the carrier wavelength $\lambda$, and therefore linearly with the carrier frequency $f_c = c/\lambda$, where $c$ is the speed of light. The result in \eqref{eq:DoF_1D} can be extended for 2D apertures and rewritten as \cite{Torkildson}
\begin{gather}\label{eq:DoF_2D}
DoF \approx \frac{A_T A_R}{R^2 \lambda^2 } + 1
\end{gather}
where $A_T$ and $A_R$ are the areas occupied by the 2D arrays at the transmitter and the receiver, respectively, so that the scaling with $f_c$ becomes quadratic. Since transmission bandwidth typically scales linearly with carrier frequency, the overall data rates can potentially scale cubically with carrier frequency.    

Advances in mmWave radio frequency integrated circuits (RFIC) in low-cost silicon semiconductor processes open up the possibility of deploying LoS MIMO at scale, for example, to boost link capacities in wireless backhaul mesh networks for urban picocells \cite{Rasekh_TWC2019}.  Consider $4\times4$ LoS MIMO with a link distance of $100\,$m. At a carrier frequency of 140 GHz,
the form factor required for a well-conditioned spatial channel is small enough to permit opportunistic deployment (e.g., on lampposts): the inter-antenna spacing for orthogonal eigenmodes is 33 cm. With QPSK modulation and $10-20\,$GHz bandwidth, we can achieve $80-160\,$Gbps uncoded data rates. Therefore, with lightweight channel coding, $100\,$Gbps becomes a feasible target. However, can we leverage the economies of scale in digital computation to realize such transceivers at reasonable cost and power consumption, using all-digital
baseband signal processing? While this is standard in modern communication receivers operating at lower bandwidths (typically below 1 GHz), as signaling bandwidths increase, 
realizing high-precision analog-to-digital converters (ADCs) is a challenge \cite{Murmann, Walden}.  Motivated by these considerations, we investigate in this paper 
whether it is possible to use all-digital processing in LoS MIMO receivers with severely quantized samples.

\noindent
{\bf Contributions:} 
We investigate design of low-precision quantizers and of spatial demultiplexing with heavily quantized observations.  \\
{\it Quantizer Design:} Rather than trying to design optimal quantizers, our first goal is to design quantizers with regular structure which approach the same Shannon limit as an unquantized system at high SNR. \\
$\bullet$ Our first result is negative.  As bandwidth increases, a particularly attractive approach is phase-only quantization: this can be implemented by passing linear combinations of the real and imaginary parts of the sample through sign detectors (one-bit ADCs), and therefore
does not require automatic gain control (see \cite{Singh_2009}). However, we prove for the $2 \times 2$ QPSK system that phase-only quantizers cannot meet the unquantized benchmark at high SNR.\\
$\bullet$ Our second result shows that amplitude-phase quantization with a relatively small number of bins does attain the unquantized benchmark.
Specifically, we prove for the $2 \times 2$ QPSK system that 2-level amplitude and 8-level phase quantization works.\\
$\bullet$ For the $4 \times 4$ system, we obtain practical guidelines and design prescriptions for quantizer design. We show via mutual information computations that per-antenna quantization into equal probability regions (which maximizes per-antenna output entropy) performs better than conventional MMSQE quantization, and that I/Q quantization performs better than amplitude/phase quantization.  In particular, we show that equal probability I/Q quantization with 2 bits per real dimension, designed using a Gaussian approximation for the received samples, achieves the unquantized benchmark at high SNR for QPSK modulation, attaining a maximum data rate of $8$ bits per channel use.\\
{\it Spatial demultiplexing:} For a $4 \times 4$ QPSK system, we investigate spatial demultiplexing with observations quantized using the 2 bit I/Q quantizer that we have designed.  
We considered well-conditioned LoS MIMO channels for which linear zero-forcing detection provides near-optimal performance with unquantized observations. Our goal is to attain uncoded error probabilities of $10^{-3}$ or better,
for which reliable communication can be obtained with lightweight, high-rate error correcting codes.\\
$\bullet$ We show that linear detection with quantized observations leads to an error floor. Since mutual information computations show that the maximum rate of 8 bits per channel use is attainable with moderate SNR penalty for the quantizer design used, we expect maximum likelihood detection not to exhibit an error floor.  We show that this is indeed the case, but the prohibitive complexity (exponential in the number
of transmitted bits) motivates design of lower-complexity spatial demultiplexing schemes.  \\
$\bullet$ We introduce the concept of {\it virtual quantization,} modeling the uncertainty created by quantization as a nuisance parameter, so that the task of estimating
the transmitted symbols can be approached using the tools of composite hypothesis testing.  We employ a Generalized Likelihood Ratio Test (GLRT) approach which leverages the efficacy of linear detection for the well-conditioned MIMO channels
considered here. A key computational advantage of the proposed approach is that, unlike maximum likelihood detection, its complexity does not scale with constellation size. In addition, we show that our proposed virtual quantization concept is resistant to the changes in channel condition due to suboptimal separation of the transmitter and the receiver (up to $\pm 20\%$ variations of the nominal range). \\
$\bullet$ While the bulk of our numerical examples are for QPSK, we also present results for 16QAM, demonstrating that our prescriptions for quantizer design and our proposed virtual quantization approach extend to larger constellations.

\textit{Notation:} Throughout the paper, random variables are denoted by capital letters and small letters are used for the specific value that the random variables take. Bold letters are used to denote vectors and matrices. $\mathop{\mathbb{E}_Z}$ denotes the expectation operator over the random variable $Z$. $|Z|$ and $\angle Z$ represent the amplitude and the phase of $Z$, respectively. $\Re(Z)$ and $\Im(Z)$ denote real and imaginary part of complex number $Z$, respectively. $\vX^\intercal$ and $\vX^\dagger$ are the transpose and Hermitian transpose of $\vX$, respectively. $\vI_n$ is the identity matrix of size $n$.

%% file: related_work.tex
\section{Related Work}

The DoF for LoS MIMO as a function of transceiver form factor and antenna placement, range and carrier frequency are by now well known \cite{Bohagen2007, Torkildson}.
It is worth contrasting the motivation for our work with a recently developed LoS MIMO system \cite{LoSMIMO_Ericsson} which employs $2.5\,$GHz bandwidth in E-band ($70-80\,$GHz carrier frequency), and achieves $100+\,$Gbps at a distance of 1.5 km using 8-fold multiplexing (spatial degrees of freedom along with dual polarization) and alphabets as large as 64QAM.  The optimal antenna separation is $1.72\,$m, requiring bulky antenna structures and careful installation.  
We envision higher frequencies and shorter ranges to reduce form factor to enable opportunistic deployment. 
The goal of our investigation of severely quantized LoS MIMO, therefore, is to examine how far we can push the paradigm of using larger available bandwidths (which limits
the precision of available ADCs) to reduce the required constellation size (which potentially enables reduction in ADC precision).
Our work also contrasts with recent efforts in the research literature based on analog-centric \cite{Sawaby_Asilomar2016, Mamandipoor, Sheldon_APS2010,Yan_2018} or hybrid analog-digital \cite{Wadhwa_2016, Khalili_ISIT2018, Raviteja_Spawc2018, Zhu_2019} processing in an attempt to sidestep the ADC bottleneck.

Since LoS MIMO is often envisioned for quasi-static links (e.g., wireless backhaul), it is natural to consider precoding with channel state information at the transmitter, as in a number of
theoretical studies \cite{Kobayashi_2007, Zhou_2014, Zhu_2019}.  Transmit precoding can also significantly reduce the dynamic range at the receiver, easing the task of
analog-to-digital conversion.  Indeed, prior studies of MIMO capacity with low-precision ADC assume
transmit precoding \cite{Mo_TSP, Khalili_ISIT2018, Khalili_2020}. Channel capacity with 1-bit ADC is studied in \cite{Mo_TSP}, which provides capacity bounds and a convex optimization based algorithm to obtain capacity-achieving constellations. In \cite{Khalili_ISIT2018}, the capacity with transmit precoding, together with hybrid analog-digital processing at the receiver, where analog linear combinations of the signals received at different antennas are quantized, is studied. In \cite{Khalili_2020}, joint transmit power and ADC allocation problem is studied for throughput maximization, which results in that using few one-bit ADCs with the adaptive threshold receiver is enough to achieve near optimal performance.

Transmit precoding leads to increased dynamic range at the transmitter, which aggravates the already difficult problem of producing power at higher frequencies, such as
the millimeter wave or THz bands.  In this paper, therefore, we explore LoS MIMO {\it without} transmit precoding, in contrast to the cited
prior work.  We assume that the receiver has ideal channel estimates.  Channel estimation with low-precision ADC is not as challenging as demodulation: \cite{Chan_Est_Dabeer} is an early example for a SISO dispersive channel, while \cite{Nossek} and \cite{Mo_2014} propose effective estimation techniques for massive MIMO with 1-bit quantization at the receive antennas.

There have been prior studies of demodulation \cite{Mezghani_2008, Mezghani_2010} based on quantized samples for MIMO systems without precoding, but these consider
Rayleigh faded channel models associated with rich scattering environments, unlike the LoS MIMO setting considered here.  The computational intractability of maximum likelihood
detection is pointed out in \cite{Mezghani_2008}, while large system analysis for suboptimal loopy belief propagation is considered in \cite{Mezghani_2010}. 

Shannon limits for an ideal SISO discrete-time additive white Gaussian noise (AWGN) channel with low-precision ADC are studied in \cite{TCOM_Madhow}. It is shown that the optimal input distribution is discrete and can be computed numerically, but standard constellations are near-optimal. Further, the use of ADCs with 2-3 bits precision results in only a small reduction in channel capacity even at moderately high SNR. Our model is perhaps the simplest possible extension of this framework to MIMO systems.

This paper builds on our preliminary results on quantizer design in an earlier conference paper \cite{ADS_UM_2019}.  We provide proofs and technical details, as well as more
detailed insights and numerical results, for quantizer design here.  The results on  spatial demultiplexing, including the proposed virtual quantization concept, are entirely new.

%% file: system_model_prob_formulation.tex
\section{System Model and Problem Formulation}\label{sec:probForm}

\begin{figure}
\vspace{-0.2cm}
\begin{center}		
\begin{subfigure}[b]{0.70\textwidth}
\includegraphics[width=\textwidth,draft=false]{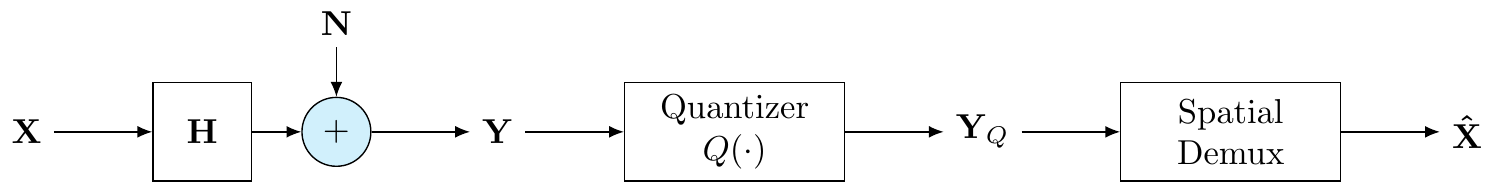}
\caption{Mathematical model for LoS MIMO communication system}\label{fig:MathModel}
\end{subfigure}
\\
\begin{subfigure}[b]{0.70\textwidth}
\includegraphics[width=\textwidth,draft=false]{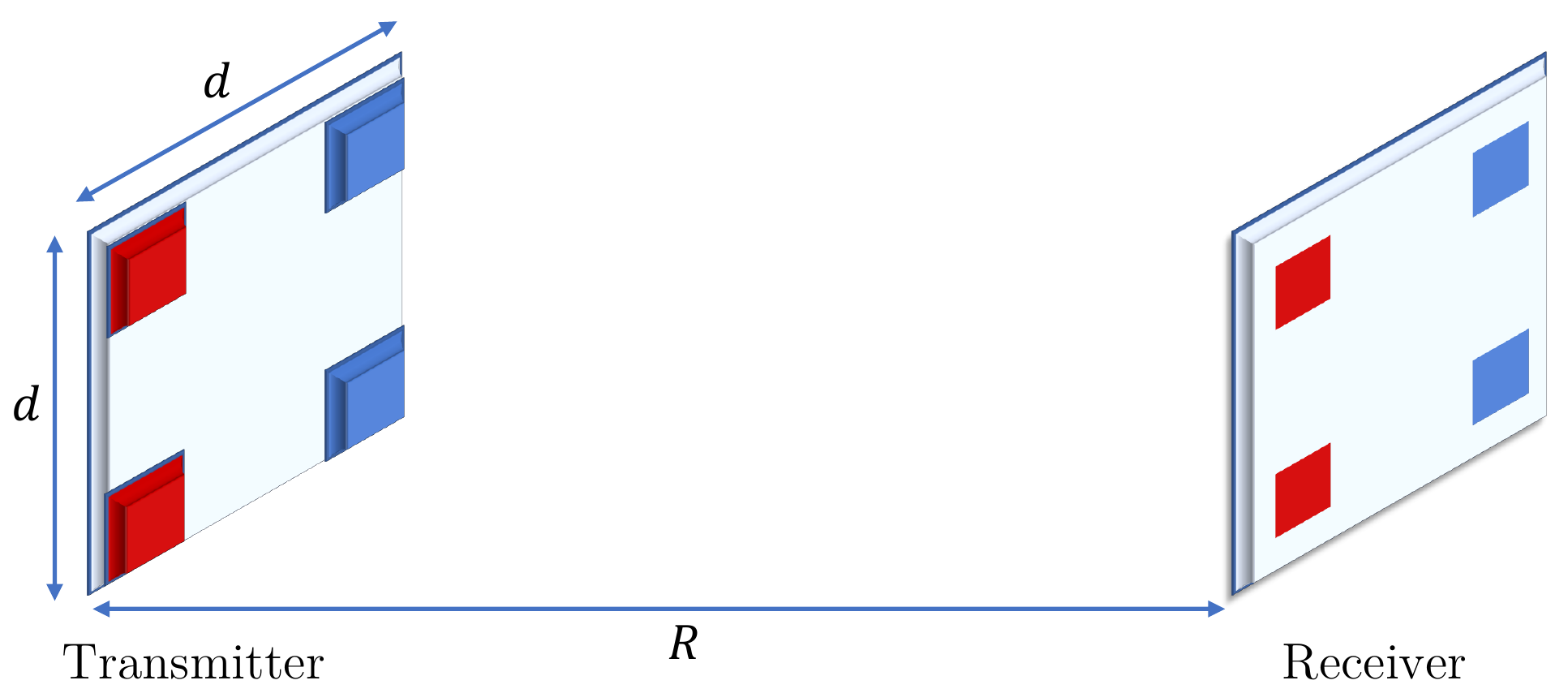}
\caption{Geometric configuration for the $4\times 4$ LoS MIMO system (with blue and red colored antennas) and the $2\times 2$ LoS MIMO system (with only red colored antennas)}\label{fig:SystemConf}
\end{subfigure}
\caption{LoS MIMO communication system model}\label{fig:SystemModel}
\end{center}
\vspace{-0.6cm}
\end{figure}

We consider a symmetric $4 \times 4$ LoS MIMO communication system, with equal inter-antenna spacings at transmitter and receiver, as shown in Fig.~\ref{fig:SystemModel}. Each transmit/receive antenna may be a fixed beam antenna \cite{LoSMIMO_Ericsson}, or an electronically steerable ``subarray'' \cite{Torkildson}, with a directive beam along the LoS, and multipath is ignored. The received signal vector $\vY \triangleq \left[Y_1 \, \cdots \, Y_{4} \right]^\intercal \in \mathbb{C}^{4 \times 1}$ is given by
\begin{gather}\label{eq:ReceivedSignal}
\vY = \vH\,\vX + \vN \,,
\end{gather}
where $\vX \triangleq \left[X_1 \, \cdots \, X_{4} \right]^\intercal \in \mathbb{C}^{4 \times 1}$ is the transmitted symbol vector, $\vH \in \mathbb{C}^{4 \times 4}$ is the normalized channel matrix (with each column normalized to unit norm), and $\vN \sim\mathcal{CN}(0, \sigma^2\,\vI_{4})$ is AWGN. Under this normalization, the SNR for the $k$th data stream is given by $SNR = \mathop{\mathbb{E}}\{ \lvert X_k \rvert^2 \}/\sigma^2$.

\noindent
{\bf Input:} We consider QPSK modulation unless otherwise stated (results for 16QAM are included in Section \ref{sec:SpatialDemultiplexing}). For QPSK modulation, $\{X_i\}_{i=1}^{4}$ are independent and identically distributed symbols taking values $\{e^{j\pi/4}, e^{j3\pi/4}, e^{j5\pi/4}, e^{j7\pi/4}\}$ with equal probability. Thus, $SNR =\mathop{\mathbb{E}}\{ \lvert X_k \rvert^2 \}/\sigma^2 = 1/\sigma^2$ where $\mathop{\mathbb{E}}\{ \lvert X_k \rvert^2 \} = 1$ for all $k\in\{1,\ldots,4\}$.

\noindent
{{\bf Channel:} 
For the pure LoS channel we consider, the elements of $\vH$ in \eqref{eq:ReceivedSignal} are calculated by employing {\it ray-tracing} in consideration of the spherical nature of the wave propagation \cite{Driessen_1999, Bohagen_2009} and the columns of $\vH$ are normalized to unit norm. Since the path loss differences among different transmit-receive antenna pairs are negligible, the normalized channel matrix for the symmetric $4 \times 4$ LoS MIMO communication system is given by \cite{Mamandipoor}
\begin{gather}\label{eq:ChannelMatrix_4x4}
\vH = \frac{1}{2} e^{-j\Phi}
\begin{bmatrix}
    1                &     e^{-j\theta}    &     e^{-j2\theta} &     e^{-j\theta} \\
    e^{-j\theta}      &     1              &     e^{-j\theta}  &     e^{-j2\theta} \\
    e^{-j2\theta}     &     e^{-j\theta}    &     1            &     e^{-j\theta}  \\
    e^{-j\theta}      &     e^{-j2\theta}   &     e^{-j\theta}  &     1            
\end{bmatrix}\,,
\end{gather}
where the random variable $\Phi$ denotes the common phase change along the path between the transmitter and the receiver, and the ``cross-over phase'' depends on the inter-antenna spacing $d$ and link distance $R$ as follows \cite{Torkildson, Garcia_2018}:
\begin{gather} \label{eq:theta}
\theta = \frac{2\pi}{\lambda}(\sqrt{R^2+ d^2} - R) \approx \frac{\pi d^2}{\lambda R} ~{\rm for}~ R \gg d
\end{gather}
where $\lambda$ denotes the carrier wavelength. We would like our quantizer designs to be robust to variations in the common phase $\Phi$, which is assumed to be uniformly distributed over $[0, 2\pi)$. 

\noindent
{\bf Quantizer:} We consider identical quantizers at each receive antenna. The quantized output of the $i$th receive antenna can be expressed as
\begin{gather}\label{eq:QuantizedOutput}
\bar{Y}_i = Q(Y_i) \,,
\end{gather}
for $i\in\{1,\ldots,4\}$. $Q(\cdot)$ in \eqref{eq:QuantizedOutput} represents the quantizer function at each receive antenna and for a given input $y$, $Q(y)$ can be characterized as
\begin{gather}\label{eq:Quantizer}
Q(y) = \tilde{y}_j\,,~\textrm{if}\;~y\in \Gamma_j\,,
\end{gather}
for $j\in\{1,\ldots,T\}$, where $\tilde{y}_j$ for $j\in\{1,\ldots,T\}$ is a design parameter and $\Gamma_1$, $\ldots$, $\Gamma_{T}$ denote the decision regions for the quantizer, with $T$ denoting the number of quantizer bins at each receive antenna.

\noindent
{\bf Spatial demultiplexer:} Based on the quantized observations; that is, $\vYQ \triangleq \left[\bar{Y}_1 \, \cdots \, \bar{Y}_{4} \right]^\intercal \in \mathbb{C}^{4 \times 1}$, the receiver performs the spatial demultiplexing and provides the estimate of the transmitted symbol as $\vXH \triangleq \left[\hat{X}_1 \, \cdots \, \hat{X}_{4} \right]^\intercal \in \mathbb{C}^{4 \times 1}$.

We begin with the $2 \times 2$ system depicted in Fig.~\ref{fig:SystemConf} with red colored antennas in order to make some fundamental theoretical observations regarding quantization. For this scheme, $\vX$ and $\vY$ in \eqref{eq:ReceivedSignal} are defined as $\vX \triangleq \left[X_1 \, X_2 \right]^\intercal \in \mathbb{C}^{2 \times 1}$ and $\vY \triangleq \left[Y_1 \, Y_2 \right]^\intercal \in \mathbb{C}^{2 \times 1}$, respectively. Also, $\vN \sim\mathcal{CN}(0, \sigma^2\,\vI_2)$. The channel matrix for this scheme corresponds to the renormalized version of the upper-left $2 \times 2$ submatrix of \eqref{eq:ChannelMatrix_4x4} and is given by
\begin{gather}\label{eq:ChannelMatrix}
\vH = \frac{1}{\sqrt{2}} e^{-j\Phi}
\begin{bmatrix}
    1                &     e^{-j\theta} \\
    e^{-j\theta}      &     1 
\end{bmatrix}\,,
\end{gather}
where $\theta \approx \frac{\pi d^2}{\lambda R}$ for $R \gg d$ as in (\ref{eq:theta}).

One possible formulation of optimal quantization is to minimize 
\begin{gather}\label{eq:Difference}
D(\vX, \vYQ, \theta) \triangleq \mathop{\mathbb{E}_\Phi}\{I(\vX; \vY \mid \Phi, \theta) - I(\vX; \vYQ \mid \Phi, \theta)\}\, 
\end{gather}
where $\vYQ \triangleq \left[\bar{Y}_1 \, \bar{Y}_2 \right]^\intercal$ and the function $I(\bar{\vX}; \bar{\vY}\mid \Phi, \theta)$ represents the mutual information between the random variables $\bar{\vX}$ and $\bar{\vY}$ for given $\Phi$ and $\theta$. Based on the data processing equality, $D(\vX, \vYQ, \theta) \geq 0$ since $\vX$, $\vY$, and $\vYQ$ form a Markov chain; that is, $\vX \rightarrow \vY \rightarrow \vYQ$. Also, $I(\vX; \vY \mid \Phi, \theta)$ in \eqref{eq:Difference}  does not depend on any parameter related to quantizer. For that reason, the problem of minimizing $D(\vX, \vYQ, \theta)$ in \eqref{eq:Difference} is equivalent to
\begin{gather}\label{eq:OptProb}
\underset{\{\Gamma_j\}_{j=1}^{T}}\max~ \mathop{\mathbb{E}_\Phi}\{I(\vX; \vYQ \mid \Phi, \theta)\}\,. 
\end{gather}
In the optimization problem in \eqref{eq:OptProb}, the mutual information between $\vX$ and $\vYQ$ must be maximized over the set of all possible quantization regions of the quantizer at the receive antennas. The number of quantization bins for the quantizer is not fixed in \eqref{eq:OptProb}, and must also be optimized.  Thus, it is difficult to solve \eqref{eq:OptProb}. Furthermore, the optimal
quantizers may correspond to irregular regions, leading to implementation difficulties. In this paper, therefore,
we opt for designing regular quantizers with the goal of ensuring that $D(\vX, \vYQ, \theta)  \rightarrow 0$ at high SNR.

%% file: 2by2System.tex
\section{Analytical Insights from $2 \times 2$ MIMO}

In this section, we provide detailed insight regarding quantization via the $2 \times 2$ system.  By analysis of a limiting noiseless regime, we prove that phase-only quantization cannot yield unique decodability, while amplitude-phase quantization can.

\subsection{Phase-only Quantization}\label{sec:phaseOnly}

The following lemma establishes a negative result for phase-only quantization.
Consider a pair of possible transmitted symbol vectors $\vX^{(1)} = (X_1^{(1)}, X_2^{(1)})$ and $\vX^{(2)} = (X_1^{(2)}, X_2^{(2)})$. The corresponding noise-free received samples are given by
$Y_1^{(i)} \triangleq e^{-j\phi}(X_1^{(i)} + e^{-j\theta} X_2^{(i)} )/\sqrt{2}$ and $Y_2^{(i)} \triangleq e^{-j\phi}(e^{-j\theta} X_1^{(i)} +  X_2^{(i)} )/\sqrt{2}$ for $i\in\{1,2\}$.
The lemma specifies choices for $\vX^{1}$ and $\vX^{2}$ for which the noise-free received samples prior to quantization
have the same phase.  Thus, these pairs cannot be distinguished based on phase-only quantization.

\textbf{Lemma 1:} 
 \textit{For $(X_1^{(1)}, X_2^{(1)}) = (e^{j\pi (2i-1)/4}, e^{j\pi(2i+1)/4})$ where $i\in\{1,\ldots, 4\}$, the following statements hold:}
\begin{enumerate}
\item[] \textit{(i) For $\theta \in (-\pi/2, \pi/2)$ and $(X_1^{(2)}, X_2^{(2)}) = (X_2^{(1)}, X_1^{(1)})$,}
\begin{align}\label{eq:EqualPhase}
\angle Y_1^{(1)} = \angle Y_2^{(1)},\,
\angle Y_1^{(2)} = \angle Y_2^{(2)},\,
\angle Y_1^{(1)} = \angle Y_1^{(2)} 
\end{align}

\item[] \textit{(ii) For $\theta \in (\pi/2, 3\pi/2)$ and $(X_1^{(2)}, X_2^{(2)}) = (X_2^{(1)}, X_1^{(1)})$,}
\begin{align}\label{eq:EqualPhase2}
\angle Y_1^{(1)} = \angle Y_2^{(1)} + \pi ,\,
\angle Y_1^{(2)} = \angle Y_2^{(2)} + \pi ,\,
\angle Y_1^{(1)} = \angle Y_1^{(2)} + \pi
\end{align}
\item[] \textit{(iii) For $\theta \in (-\pi/2, \pi/2)$ and $(X_1^{(2)}, X_2^{(2)}) = (e^{j\pi}X_2^{(1)}, e^{j\pi}X_1^{(1)})$,}
\begin{align}\label{eq:EqualPhase3}
\angle Y_1^{(1)} = \angle Y_2^{(1)} ,\,
\angle Y_1^{(2)} = \angle Y_2^{(2)} ,\,
\angle Y_1^{(1)} = \angle Y_1^{(2)} + \pi
\end{align}

\item[] \textit{(iv) For $\theta \in (\pi/2, 3\pi/2)$ and $(X_1^{(2)}, X_2^{(2)}) = (e^{j\pi}X_2^{(1)}, e^{j\pi}X_1^{(1)})$,}
\begin{align}\label{eq:EqualPhase4}
\angle Y_1^{(1)} = \angle Y_2^{(1)} + \pi ,\,
\angle Y_1^{(2)} = \angle Y_2^{(2)} + \pi ,\,
\angle Y_1^{(1)} = \angle Y_1^{(2)}
\end{align}
\end{enumerate}

\indent\indent\textit{Proof:} The result in the lemma can simply be shown by using Euler's formula and Pythagorean trigonometric identity.

\hfill $\blacksquare$

This results in the following proposition stating that phase-only quantization cannot achieve the unquantized benchmark. 

\textbf{Proposition 1:} \textit{For any phase-only quantization scheme with any number of bins, $D(\vX, \vYQ, \theta) > 0$ for all $\theta \in [0, 2\pi)$ as $\sigma \to 0$.}

\indent\indent\textit{Proof:} In order to show that $D(\vX, \vYQ, \theta) > 0$ for all $\theta \in [0, 2\pi)$ as $\sigma \to 0$, $\mathop{\mathbb{E}_\Phi}\{I(\vX; \vYQ \mid \Phi, \theta)\} < 4$ should be proved for $\theta \in [0, 2\pi)$ as $\sigma \to 0$ since $ \mathop{\mathbb{E}_\Phi}\{I(\vX; \vY \mid \Phi, \theta) \to 4$ as $\sigma \to 0$. Before proving that, first, it is shown that $p(\vX = \vx \mid \vYQ = \vyQ, \Phi = \phi, \theta)$ satisfies for some $\vYQ = \vyQ$ that $0 < p(\vX = \vx \mid \vYQ = \vyQ, \Phi = \phi, \theta)< 1$ as $\sigma \to 0$. Based on the statement in Lemma~1, it can be stated that the noise-free outputs corresponding to inputs $(X_1^{(1)}, X_2^{(1)}) = (e^{j\pi (2i-1)/4}, e^{j\pi(2i+1)/4})$ for $i\in\{1,\ldots, 4\}$ and $(X_1^{(2)}, X_2^{(2)}) = (X_2^{(1)}, X_1^{(1)})$ have the same phase and fall in the same quantization bin for $\theta \in (-\pi/2, \pi/2)$. Similarly, for $(X_1^{(1)}, X_2^{(1)}) = (e^{j\pi (2i-1)/4}, e^{j\pi(2i+1)/4})$ for $i\in\{1,\ldots, 4\}$ and $(X_1^{(2)}, X_2^{(2)}) = (e^{j\pi}X_2^{(1)}, e^{j\pi}X_1^{(1)})$, the outputs without additive noise stay in the same bin of any given phase-only quantization mapping since they have the same phase for $\theta \in (\pi/2, 3\pi/2)$. In addition, for $\theta = \pi/2$ and $\theta = 3\pi/2$, the amplitude of one of the noise-free outputs is zero and the same ambiguity occurs for those cases as well. Without loss of generality, say that those noiseless outputs (i.e., the outputs with additive Gaussian noise as $\sigma \to 0$) after quantization is $ \vYQ = \bvyQ$ for the inputs $\vX = \vx_1$ and $\vX = \vx_2$. Then, the following statements hold for $i\in\{1, 2\}$ based on Bayes' theorem:
\begin{align}\label{eq:CondProb}
0 < p(\vx_i \mid \vYQ = \bvyQ, \phi, \theta) = \frac{p(\vYQ = \bvyQ \mid \vx_{i}, \phi, \theta)}{\sum_{\bvx}~p(\vYQ = \bvyQ \mid \bvx, \phi, \theta)}  < 1
\end{align}
since $p(\vYQ = \bvyQ \mid \vx_{1}, \phi, \theta) \to 1$ and $p(\vYQ = \bvyQ \mid \vx_{2}, \phi, \theta) \to 1$ as $\sigma \to 0$. Then, as $\sigma \to 0$, $H(\vX \mid \vYQ, \Phi = \phi, \theta)>0$ for all $\theta \in [0, 2\pi)$ and consequently $I(\vX; \vYQ\mid \Phi = \phi, \theta)<4$ for all $\phi\in [0, 2\pi)$ based on the definition of mutual information and $\mathop{\mathbb{E}_\Phi}\{I(\vX; \vYQ \mid \Phi, \theta)\} < 4$ is satisfied for all $\theta \in [0, 2\pi)$.
\hfill $\blacksquare$

While the unquantized benchmark cannot be achieved, it is still of interest to ask how many phase quantization bins are enough to reach the high-SNR asymptote for phase-only quantization.
We now establish that, for our system, 8 phase quantization bins suffice. We begin with the following lemma. 

\textbf{Lemma 2:} \textit{For any possible $(X_1^{(1)}, X_2^{(1)})$ and $(X_1^{(2)}, X_2^{(2)})$ input pairs, $\angle Y_1^{(1)} - \angle Y_1^{(2)} = 0~\pmod{\pi/4}$ and $\angle Y_2^{(1)} - \angle Y_2^{(2)} = 0~\pmod{\pi/4}$, where $Y_1^{(i)}$ and $Y_2^{(i)}$ are as defined in Lemma~1. Also, $\angle Y_1^{(1)} - \angle Y_1^{(2)}$ and $\angle Y_2^{(1)} - \angle Y_2^{(2)}$ can take $8$ different values.}

\indent\indent\textit{Proof:} By using $\arctan(x) - \arctan(y) = \arctan(\frac{x-y}{1+xy})$ and Euler's formula, the proof is straightforward.
\hfill $\blacksquare$

Based on Lemma~2, we can derive the following proposition stating that 8 phase quantization bins suffice. 

\textbf{Proposition 2:} \textit{As $\sigma \to 0$, any phase-only quantization schemes with more than 8 regions cannot achieve higher data rate than phase-only quantization scheme with 8 equally partitioned sectors.}

\indent\indent\textit{Proof:} Consider a phase-only quantization scheme having more than 8 bins and let $L>8$ denote the number of bins of that scheme. Lemma~2 implies that the noise-free outputs (i.e., the outputs as $\sigma \to 0$) of the all possible inputs can take 8 different phase values for given $\phi$ and $\theta$. Then, based on the pigeonhole principle, at least $L-8$ bins of the phase-only quantizer do not contain any outputs for given $\phi$ and $\theta$ as $\sigma \to 0$. In other words, none of the outputs corresponding to all possible inputs fall into those bins as $\sigma \to 0$. Let $\mathcal{\bar{S}}$ denote the index set of those empty bins. Then, define a new set, $\mathcal{S}_E$ as $\mathcal{S}_E = \{\vy = [i,j] \mid ( i\in \mathcal{\bar{S}} \land j \in \{1,\ldots, L\})\, \lor\, (j\in \mathcal{\bar{S}} \land i \in \{1,\ldots, L\}) \}$. It can be stated that $p(\vYQ = \vyQ) = 0$ for all $\vyQ \in \mathcal{S}_E$ as $\sigma \to 0$. Now, two cases should be analyzed separately. First, if exactly $L-8$ bins of the phase-only quantizer are empty as $\sigma \to 0$; then, any two different noise-free outputs having different phases cannot be in the same bin due to the result in Lemma~2, which also holds for the phase-only quantization scheme with equally divided 8 regions. Let $\mathcal{S}^L_Q$ and $\mathcal{S}_Q$ denote the sets of all possible quantized outputs for the quantization scheme having $L>8$ bins and 8 bins, respectively. There exists a one-to-one correspondence between $\mathcal{S}^L_Q\backslash \mathcal{S}_E$ and $\mathcal{S}_Q$ and if $\vk \in \mathcal{S}^L_Q\backslash \mathcal{S}_E$ and $\vl \in \mathcal{S}_Q$ are the paired elements, it is stated that the input producing quantized output $\vk$ in the quantization scheme with more than 8 bins produces $\vl$ in the quantization scheme with equally partitioned 8 regions as $\sigma \to 0$. Thus,
\begin{align}\label{eq:DiscardEmptySet}
\sum_{\vyQ \in\mathcal{S}^L_Q} H(\vX \mid \vyQ) p(\vyQ) &= \sum_{\vyQ \in\mathcal{S}^L_Q\backslash \mathcal{S}_E} H(\vX \mid \vyQ) p(\vyQ) \\\label{eq:SameMutu}
& = \sum_{\vyQ \in\mathcal{S}_Q} H(\vX \mid \vyQ) p(\vyQ)\,,
\end{align}
where \eqref{eq:DiscardEmptySet} is due to $p(\vYQ = \vyQ) = 0$ for all $\vyQ \in \mathcal{S}_E$. Therefore, both of the schemes achieve the same data rate if $L-8$ bins of the phase-only quantizer are empty as $\sigma \to 0$. Next, consider the case that more than $L-8$ bins are empty. Since the noise-free outputs can have 8 different phase values for given $\phi$ and $\theta$, some of those outputs having different phases are in the same bin, which is not a possible case for the phase-only quantization scheme with equally sized regions. For that reason, it can be calculated that $H(\vX \mid \vYQ^{(1)}, \Phi = \phi, \theta) - H(\vX \mid \vYQ^{(2)}, \Phi = \phi, \theta) \geq 0$ as $\sigma \to 0$, where $\vYQ^{(1)}$ and $\vYQ^{(2)}$ denote the quantized outputs under the quantization schemes with more than 8 regions and equally partitioned exactly 8 regions, respectively. Therefore, $I(\vX; \vYQ^{(1)} \mid \Phi = \phi, \theta) \leq I(\vX; \vYQ^{(2)} \mid \Phi = \phi, \theta)$.
\hfill $\blacksquare$

\subsection{Amplitude-Phase Quantization}\label{sec:AmpPhaseQuant}

For $K$-ary amplitude and $M$-ary phase quantization, the quantization set of $(m+M(k-1))$th-bin of a quantizer can be written as
\begin{multline}\label{eq:Amp_Quant_Region}
\bar{\Gamma}_{m+M(k-1)} = \{ \bvY \mid A_{k-1} \leq |\bvY| < A_{k} \,, \frac{2\pi}{M}(m-1) \leq \angle \bvY < \frac{2\pi}{M}m \}\,,
\end{multline}
for $m\in\{1,\ldots, M\}$ and $k\in\{1,\ldots, K\}$, where $A_1,\ldots,A_{K-1}$ are the amplitude thresholds (we set $A_0 = 0$ and $A_K = \infty$ to maintain a unified notation across quantization bins).

The following proposition states that $K=2$ and $M=8$ suffices to attain the unquantized benchmark.

\textbf{Proposition 3:} \textit{As $\sigma \to 0$, circularly symmetric quantization with $2$-level amplitude and $8$-level phase quantization
attains $D(\vX, \vYQ, \theta) \rightarrow 0$ for $\theta \in [0, 2\pi)$.}

\indent\indent\textit{Proof:} Proposition~1 is based on the observation that the outputs of some input pairs have the same phase at both of the antennas as $\sigma \to 0$, so that those outputs cannot be differentiated by employing any phase-only quantization scheme. On the other hand, the proof of Proposition~2 shows that a phase-only scheme with equally partitioned $8$ regions can distinguish noise-free outputs having two different phases, due to the result in Lemma~2. In this proof, the aim is to show that considering a 2-level amplitude quantization together with phase quantization resolves the ambiguities leading to the result in Proposition~1. First, it can be shown that only the outputs corresponding to the input pairs discussed in Lemma~1 cannot be distinguished via phase-only scheme having equally partitioned $8$ regions. For that reason, consider the input pairs in Lemma~1. For $(X_1^{(1)}, X_2^{(1)}) = (e^{j\pi (2i-1)/4}, e^{j\pi(2i+1)/4})$ and $(X_1^{(2)}, X_2^{(2)}) = (X_2^{(1)}, X_1^{(1)})$, where $i\in\{1,\ldots, 4\}$, $|Y_1^{(2)}| < 1 < |Y_1^{(1)}|$ and $|Y_2^{(1)}| < 1 < |Y_2^{(2)}|$ for $\theta \in (0, \pi/2]$, $|Y_1^{(1)}| = |Y_1^{(2)}| = 1$ and $|Y_2^{(1)}| = |Y_2^{(2)}| = 1$ for $\theta = 0$, and $|Y_1^{(1)}| < 1 < |Y_1^{(2)}|$ and $|Y_2^{(2)}| < 1 < |Y_2^{(1)}|$ for $\theta \in [-\pi/2, 0)$. Due to the symmetry, the same approach can be applied for other input pairs (i.e., $(X_1^{(1)}, X_2^{(1)}) = (e^{j\pi (2i-1)/4}, e^{j\pi(2i+1)/4})$ and $(X_1^{(2)}, X_2^{(2)}) = (e^{j\pi}X_2^{(1)}, e^{j\pi}X_1^{(1)})$ for $i\in\{1,\ldots, 4\}$) when $\theta \in (\pi/2, 3\pi/2)$. Since the amplitude of the outputs does not depend on $\Phi = \phi$ and a circularly symmetric quantization scheme is employed, a phase quantization scheme including a 2-level amplitude quantization with $A_0 = 0$, $A_1 = 1$, and $A_2 = \infty$ resolves the ambiguity between those outputs. It is easy to now conclude that $D(\vX, \vYQ, \theta) \to 0$ for $\theta \in [0, 2\pi)$ as $\sigma \to 0$.
\hfill $\blacksquare$

\subsection{Numerical Results}\label{sec:NumRes_2by2_Quantization}

In this section, numerical examples are provided to illustrate the theoretical results. We first illustrate the statements in the lemmas and the propositions
via example noise-free outputs prior to quantization.  We then compute and compare Shannon limits for different quantization schemes.

\begin{figure}
\vspace{-0.2cm}
\begin{center}
\includegraphics[width=0.70\columnwidth,draft=false]{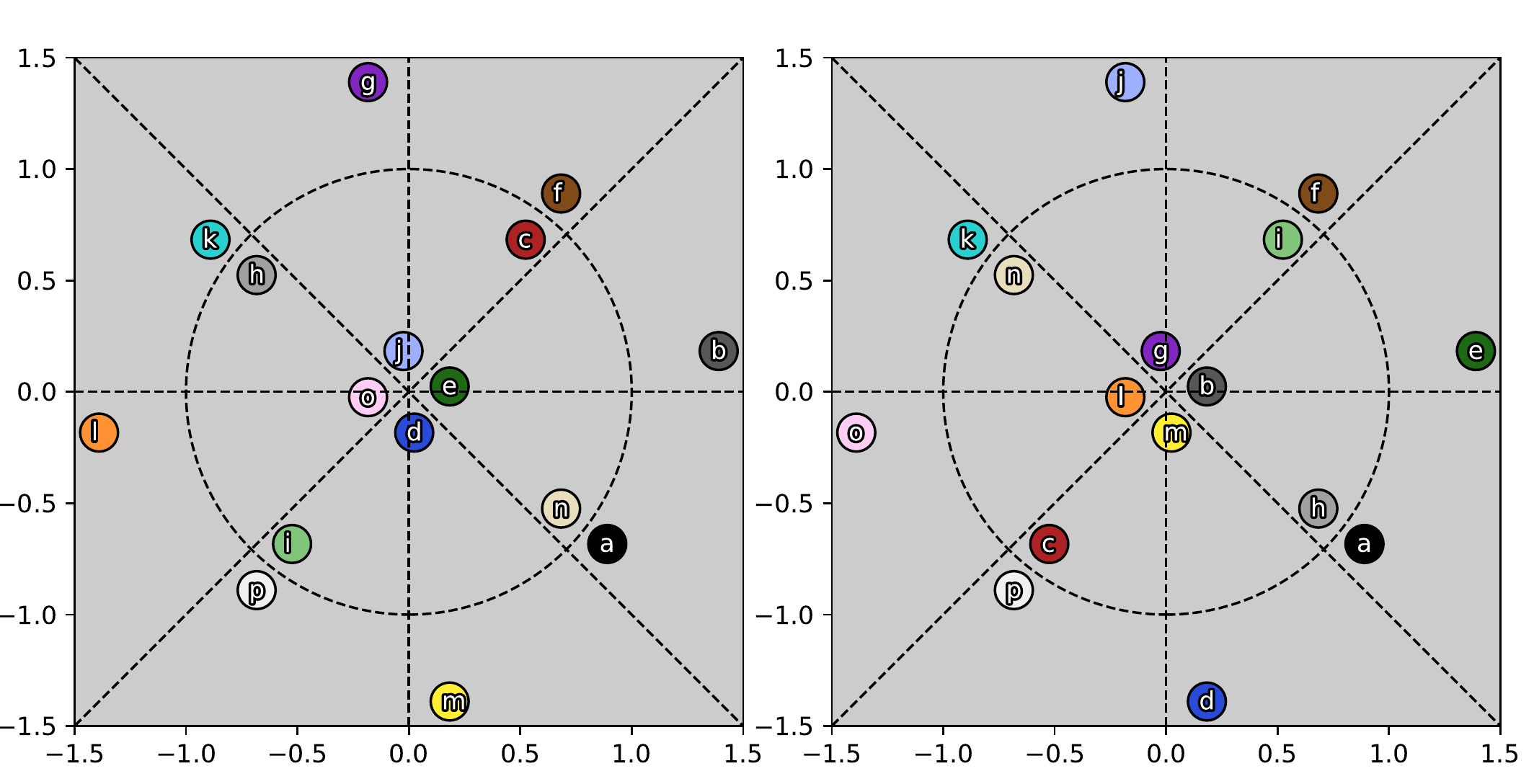}
\vspace{-0.3cm}
\caption{All possible noise-free outputs before the quantization, $e^{-j\phi}(X_1 + e^{-j\theta} X_2)/\sqrt{2}$ (Left) and $e^{-j\phi}(e^{-j\theta} X_1  + X_2 )/\sqrt{2}$ (Right), for $\theta = 5\pi/12 $ and $\phi = \pi/4$.}\label{fig:Cons_theta_75_phi_45}
\end{center}
\vspace{-0.6cm}
\end{figure}

\begin{figure}
\vspace{-0.2cm}
\begin{center}
\includegraphics[width=0.70\columnwidth,draft=false]{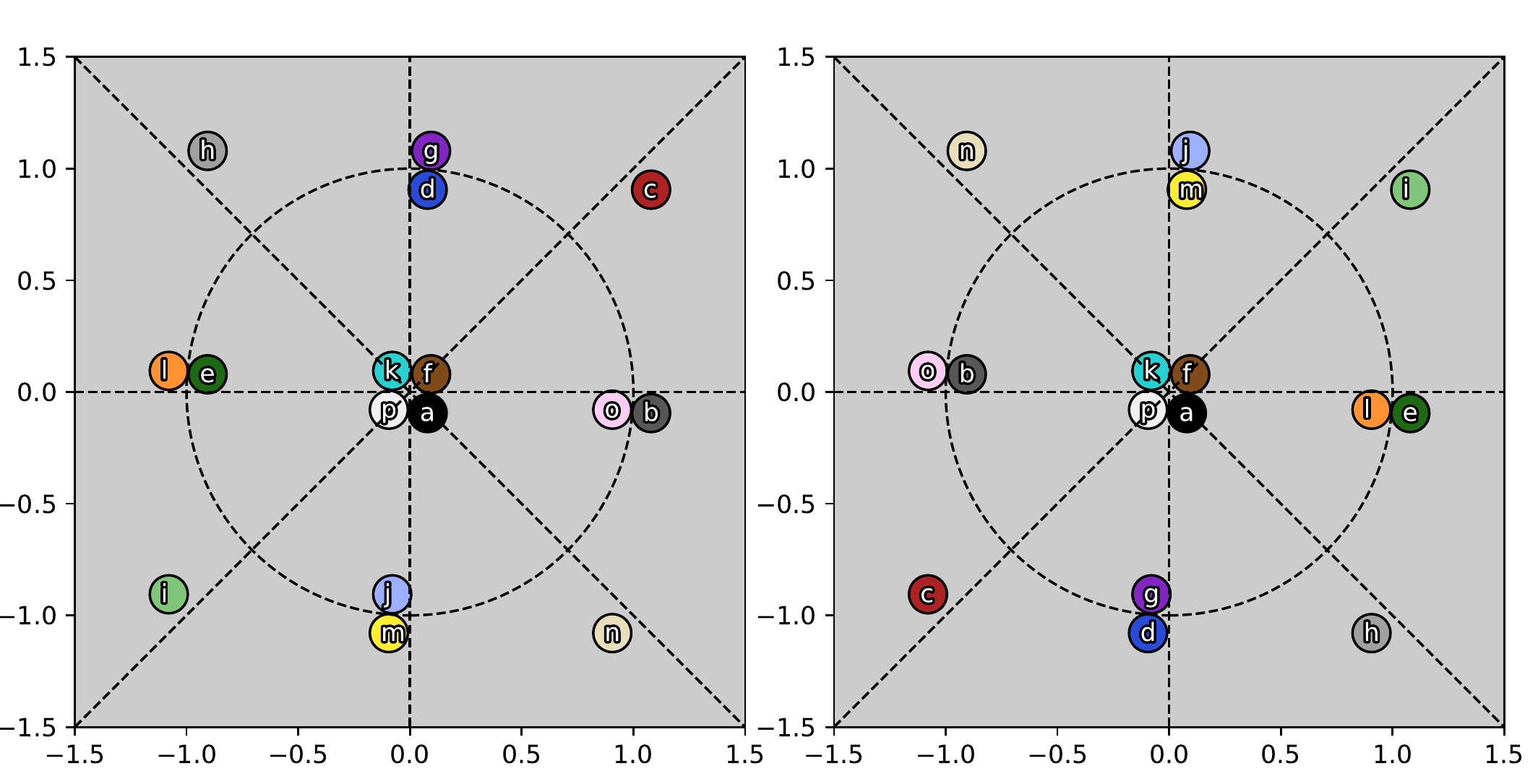}
\vspace{-0.3cm}
\caption{All possible noise-free outputs before the quantization, $e^{-j\phi}(X_1 + e^{-j\theta} X_2)/\sqrt{2}$ (Left) and $e^{-j\phi}(e^{-j\theta} X_1  + X_2 )/\sqrt{2}$ (Right), for $\theta = 17\pi/18 $ and $\phi = \pi/18$.}\label{fig:Cons_theta_10_phi_10}
\end{center}
\vspace{-0.6cm}
\end{figure}

We illustrate the geometry behind the proofs by presenting noiseless outputs prior to quantization for a well-conditioned and a poorly conditioned channel in Fig.~\ref{fig:Cons_theta_75_phi_45} and Fig.~\ref{fig:Cons_theta_10_phi_10}, respectively. We see that some output pairs (e.g., $(b, e)$, $(d, m)$, $(g, j)$ and $(l, o)$ in Fig.~\ref{fig:Cons_theta_75_phi_45} and $(b, o)$, $(d, g)$, $(e, l)$ and $(j, m)$ in Fig.~\ref{fig:Cons_theta_10_phi_10}) have the same phase at {\it both} receive antennas, and hence cannot be distinguished based on phase-only quantization, as stated in Lemma~1. On the other hand, the other outputs can indeed be distinguished based on phase-only quantization. In addition, for given $\theta$ and $\phi$, the phase of noise-free outputs can have $8$ different values, and two different outputs having two different phase values cannot be in the same bin for phase-only quantization with 8 equal sectors. This is the intuitive basis for Proposition~2. Lastly, noise-free output pairs having the same phase at both receive antennas, such as $(b, e)$ in Fig.~\ref{fig:Cons_theta_75_phi_45} can be separated by employing an amplitude quantization scheme with 2 regions as illustrated in Fig.~\ref{fig:Cons_theta_75_phi_45} and Fig.~\ref{fig:Cons_theta_10_phi_10}. This is the intuition behind Proposition~3.

\begin{figure}
\vspace{-0.2cm}
\begin{center}
\begin{subfigure}[b]{0.48\textwidth}
\includegraphics[width=\textwidth,draft=false]{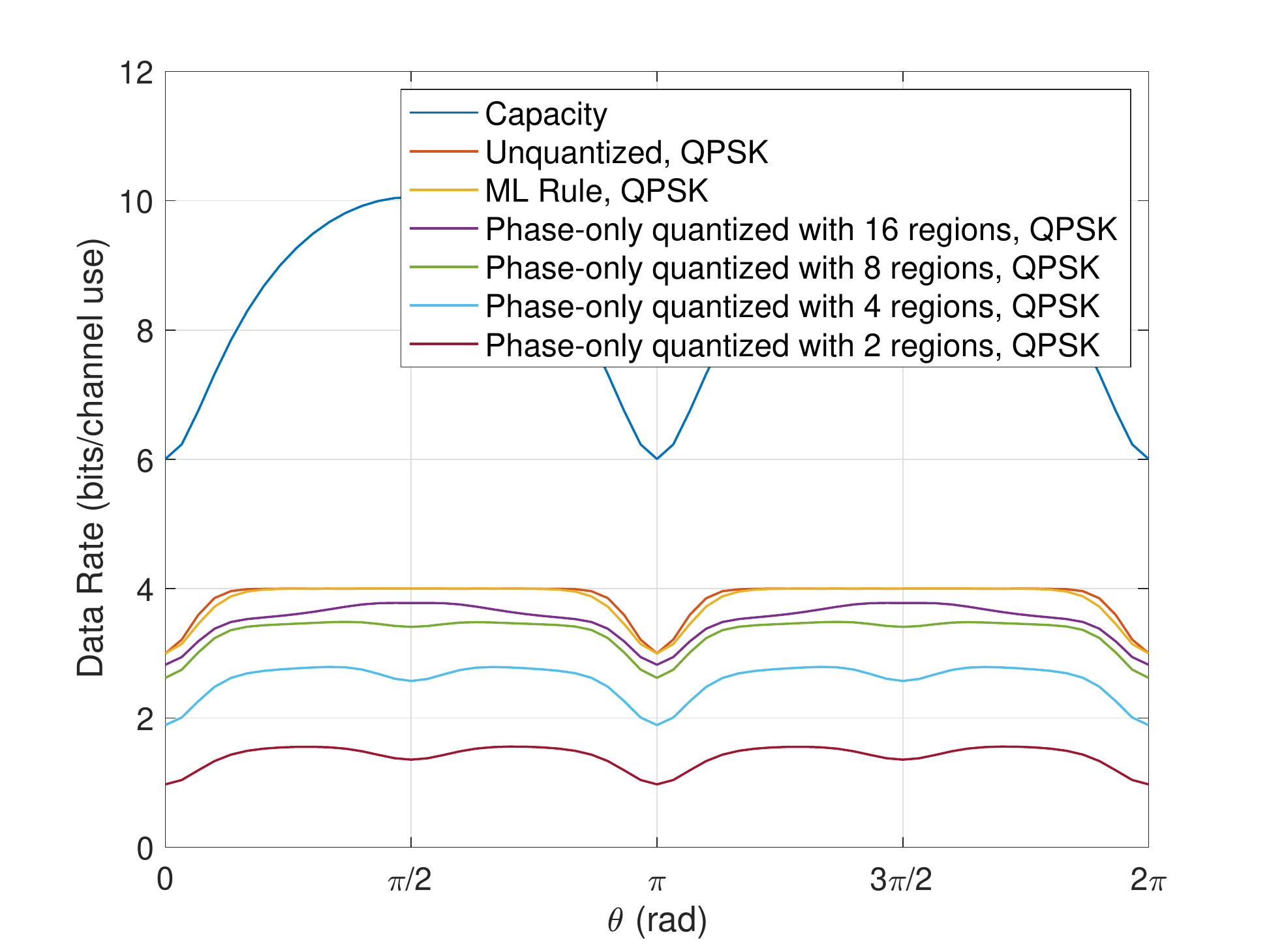}
\caption{}\label{fig:Phase_Only_Cap_VS_Theta}
\end{subfigure}
\begin{subfigure}[b]{0.48\textwidth}
\includegraphics[width=\textwidth,draft=false]{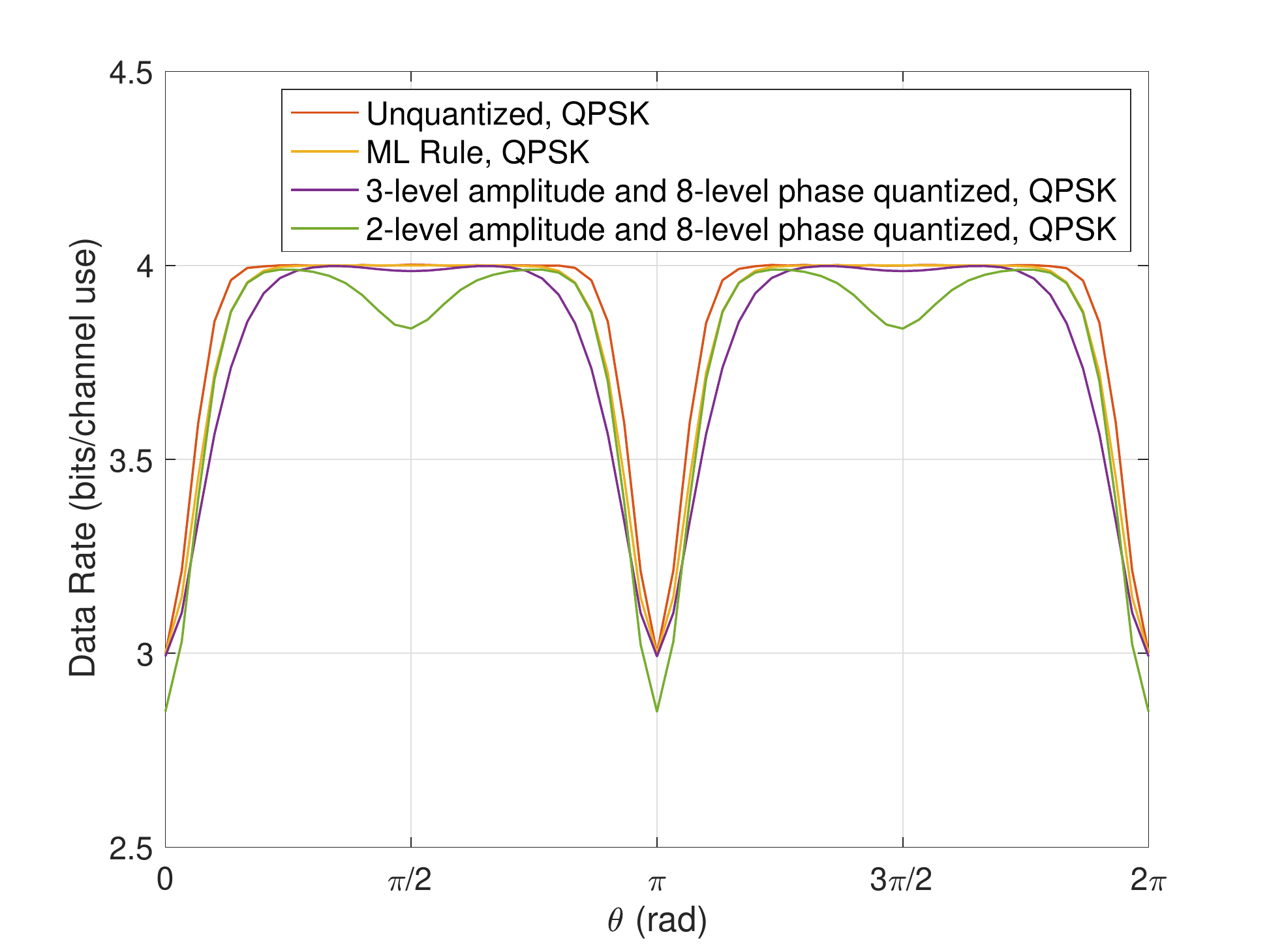}
\caption{}\label{fig:Amp_Phase_Cap_VS_Theta}
\end{subfigure}
\caption{Data rate versus $\theta$ for various scenarios including (a) the phase-only quantization schemes and (b) the amplitude and phase quantization schemes with different number of regions, where SNR is $15\,$dB.}
\end{center}
\vspace{-0.6cm}
\end{figure}

\begin{figure}
\vspace{-0.2cm}
\begin{center}
\begin{subfigure}[b]{0.48\textwidth}
\includegraphics[width=\columnwidth,draft=false]{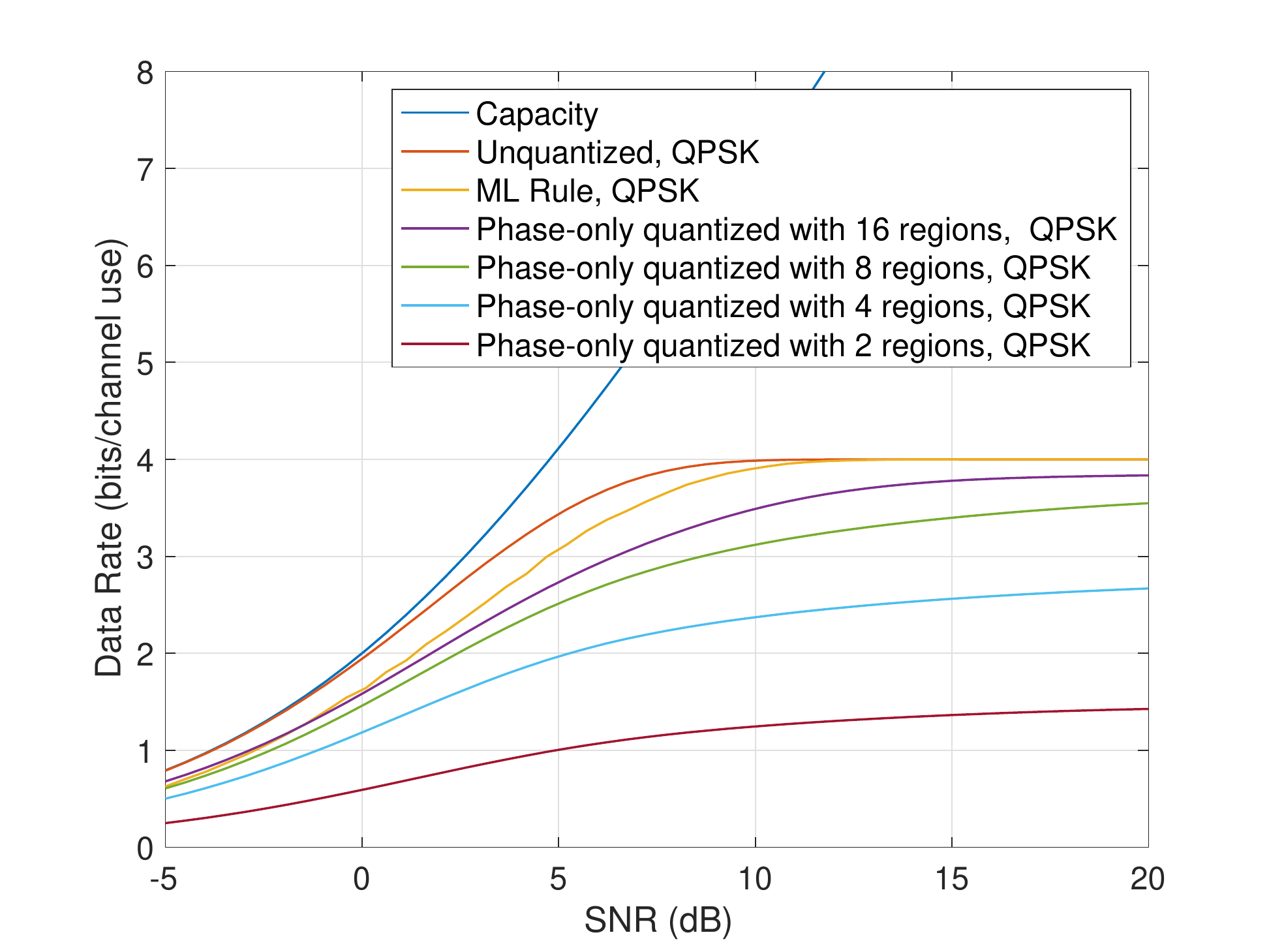}
\caption{}\label{fig:Phase_Only_Cap_VS_SNR}
\end{subfigure}
\begin{subfigure}[b]{0.48\textwidth}
\includegraphics[width=\columnwidth,draft=false]{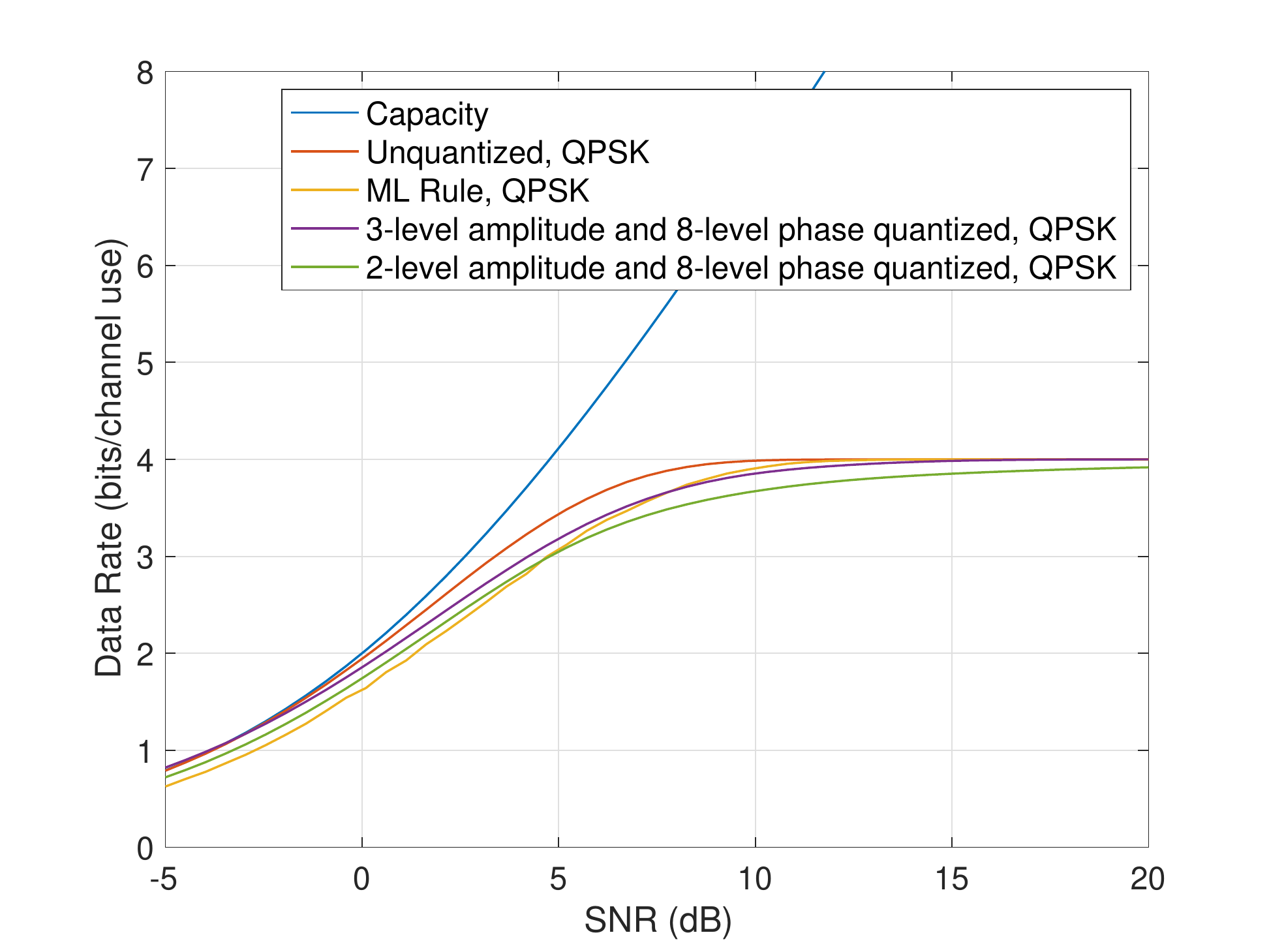}
\caption{}\label{fig:Amp_Phase_Cap_VS_SNR}
\end{subfigure}
\caption{Data rate versus SNR for various scenarios including (a) the phase-only quantization schemes and (b) the amplitude and phase quantization schemes with different number of regions, where $\theta = \pi/2$.}
\end{center}
\vspace{-0.6cm}
\end{figure}

Next, we plot the data rate (mutual information) attained by different quantization schemes.  Two benchmarks are considered: an unquantized system, and a quantizer based on Voronoi regions separating the outputs at each antenna. We may view the latter as an ML decision rule at each antenna, where input-pairs that fall on top of each other are interpreted as a single point, and it is easy to see that it attains the unquantized benchmark at high SNR. However, it depends on $\theta$ and $\Phi$, and is an irregular quantizer, which is unattractive in practice.

For a $2\times2$ MIMO system, Fig.~\ref{fig:Phase_Only_Cap_VS_Theta} and Fig.~\ref{fig:Amp_Phase_Cap_VS_Theta} plot data rate versus $\theta \in [0, 2\pi)$ at 15 dB SNR for phase-only and amplitude-phase quantization, respectively. Similarly, Fig.~\ref{fig:Phase_Only_Cap_VS_SNR} and Fig.~\ref{fig:Amp_Phase_Cap_VS_SNR} plot data rates versus SNR, fixing $\theta = \pi/2$ (the best conditioned channel). For 2-level amplitude and 8-level phase quantization, the amplitude threshold is set to $A_1 = 1$, whereas the thresholds are $A_1 = 0.75$ and $A_2 = 1.25$ for 3-level amplitude and 8-level phase quantization. The plots illustrate the trends predicted by our theoretical results: phase-only quantization does not attain the unquantized or ML benchmarks, while amplitude-phase quantization does attain these at high enough SNR. However, the performance at moderate SNR can benefit from a larger number of quantization
bins than those indicated by high-SNR asymptotics. 
For example, while 8 phase quantization bins are as good as any other phase-only quantization scheme asymptotically, using 16 quantization bins does provide better performance at moderate SNRs (Fig.~\ref{fig:Phase_Only_Cap_VS_Theta} and Fig.~\ref{fig:Phase_Only_Cap_VS_SNR}).   Similarly, while 2-level amplitude quantization suffices, there is a gain at moderate SNRs with 3-level quantization (Fig.~\ref{fig:Amp_Phase_Cap_VS_Theta} and Fig.~\ref{fig:Amp_Phase_Cap_VS_SNR}). In particular, Fig.~\ref{fig:Amp_Phase_Cap_VS_SNR} shows that for a well-conditioned channel, while 2-level amplitude quantization attains the unquantized benchmark at high enough SNR, 3-level amplitude quantization has a significant advantage at moderate SNRs, reaching unquantized performance at around 12.5 dB.

Armed with these insights, we consider quantizer design for a $4 \times 4$ system in the next section.

%% file: 4by4System.tex
\section{Quantization for $4\times4$ LoS MIMO}\label{sec:Quantizationfor4x4}

In this section, we investigate design of regular quantizers for a $4\times4$ LoS MIMO system in which the transmit and receive antennas are 
configured in a two-dimensional (2D) planar array as in Fig.~\ref{fig:SystemConf}.

\subsection{Quantizer Design}\label{sec:QuantizerDesignfor4x4}

We seek to design regular quantizers which are identical for each receive antenna.
For per-stream QPSK modulation, there are $4^4$ possible noise-free values for the received sample at each antenna, and detailed analysis as in the $2 \times 2$ MIMO
system is no longer feasible.  However, as we shall see, a relatively simple approximation for the distribution of the received samples provides an effective approach
for quantizer design.

The distribution functions of $\{Y_i\}_{i=1}^{4}$ can be expressed as
\begin{align}\label{eq:Distribution_Y}
f_{Y_i}(y_i) = \int_0^{2\pi} \frac{1}{\left|\mathcal{S}\right|^4} \mathop{\sum_{\{x_i\}_{i=1}^{4}}}_{x_i \in \mathcal{S}} f(y_i \mid \phi, \{X_i=x_i\}_{i=1}^{4})\,p(\phi)\,d\phi
\end{align}
for all $i\in\{1,\ldots, 4\}$ with
\begin{align}\label{eq:Distribution_cond_Y}
f(y_i \mid \phi, \{X_i=x_i\}_{i=1}^{4})= \frac{1}{\pi \sigma^2}e^{-\frac{\|y_i - (\vH \vx)_i\|^2}{\sigma^2}}
\end{align}
where $\mathcal{S} = \{e^{j\pi/4}, e^{j3\pi/4}, e^{j5\pi/4}, e^{j7\pi/4}\}$, $p(\phi) = 1/(2\pi)$ by the assumption, $\vx \triangleq \left[x_1 \, x_2 \, x_3 \, x_4 \right]^\intercal$, $\vH$ is as in \eqref{eq:ChannelMatrix_4x4} with $\Phi = \phi$ and $\theta$, and $(\vH \vx)_i$ selects the $i$th element of $\vH \vx$.

Due to the symmetry, it is clear that the outputs before quantization (i.e., $\{Y_i\}_{i=1}^{4}$) have the same probability density function. Hence, without loss of generality, we focus on one of the outputs before quantization (e.g., say $Y_1$) to design the corresponding quantizer and employ the same quantizer for all outputs. As seen in \eqref{eq:Distribution_Y}, $Y_1$ has a complex and intractable distribution: conditioned on the common phase $\phi$, it is a mixture of $4^4$ Gaussians, and this conditional density then needs to be averaged over the continuum
$[0, 2 \pi )$ of values taken by $\phi$. We therefore approximate this distribution by
a circularly-symmetric complex Gaussian distribution, $\tilde{Y} \sim\mathcal{CN}(\tilde{\mu}, \tilde{\sigma}^2)$, with parameters chosen to minimize the Kullback-Leibler (KL) divergence between the distributions of $Y_1$ and $\tilde{Y}$, which is given by
\begin{align}\label{eq:KL_Divergence}
D_{KL} \left( Y_1\, \middle\|\, \tilde{Y} \right) = \int_{y \in \mathbb{C}} f_{Y_1}(y) \log\left(\frac{f_{Y_1}(y)}{f_{\tilde{Y}}(y)}\right)\,dy\,.
\end{align}
The optimal $\tilde{Y}$ that minimizes the KL divergence in \eqref{eq:KL_Divergence} can be found by \emph{moment matching} \cite{Rasmussen_2004}: the first and second moments of $\tilde{Y}$ and $Y_1$ are matched to obtain the optimal Gaussian approximation. Therefore, $\tilde{\mu}$ and $\tilde{\sigma}^2$ can be calculated, respectively, as
\begin{align}\label{eq:meanmatching}
\tilde{\mu} = \mathop{\mathbb{E}}\{Y_1\} = 0
\end{align}
and
\begin{align}\label{eq:variancematching}
\tilde{\sigma}^2 &= \mathop{\mathbb{E}}\{Y_1\,Y_{1}^{\dagger}\}\\
&= \sum_{i=1}^{4} \frac{1}{4}\mathop{\mathbb{E}}\{X_{i}\,X_{i}^{\dagger}\} + \mathop{\mathbb{E}}\{N_{1}\,N_{1}^{\dagger}\}\\
&= 1 + \sigma^2
\end{align}
where $\mathop{\mathbb{E}}\{X_{i}\,X_{i}^{\dagger}\} = 1$ for all $i\in\{1,\ldots, 4\}$ by definition. As a result, we consider the complex Gaussian approximation with $\tilde{\mu} = 0$ and $\tilde{\sigma}^2 = 1 + \sigma^2$ to design quantizers for $4\times4$ LoS MIMO.

\begin{figure}
\vspace{-0.2cm}
\begin{center}
\includegraphics[width=0.70\columnwidth,draft=false]{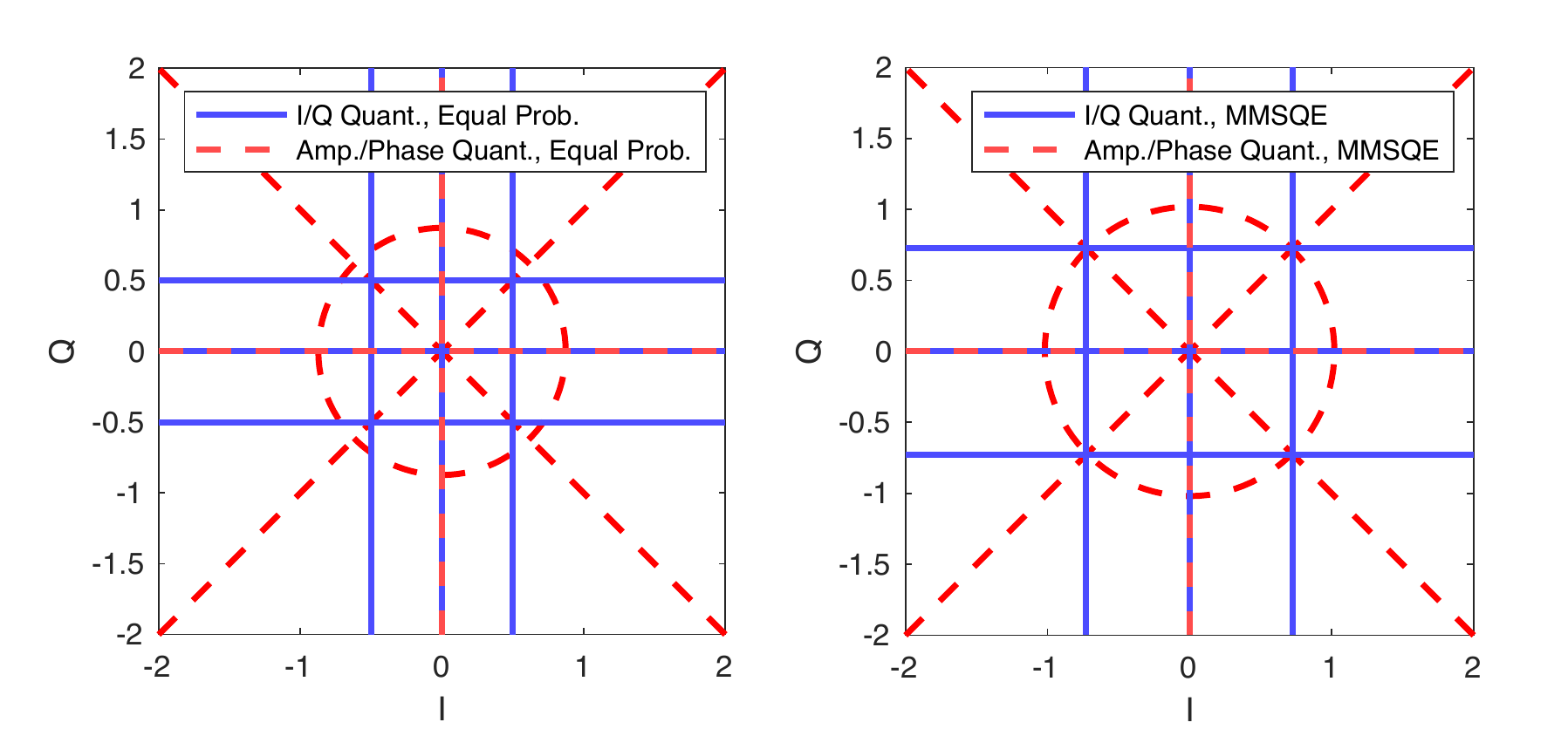}
\vspace{-0.3cm}
\caption{Quantization schemes for $4\times4$ LoS MIMO at 10 dB SNR}\label{fig:Quantization_4x4}
\end{center}
\vspace{-0.6cm}
\end{figure}

We consider two regular quantization schemes: I/Q quantization and amplitude/phase quantization.
\begin{itemize}
  \item \textbf{I/Q quantization:} For I/Q quantization scheme with $S^2$ regions, the quantization set of $(j+S(i-1))$th-bin of a quantizer can be written as
\begin{multline}\label{eq:I_Q_Region}
\bar{\Gamma}_{j+S(i-1)} = \{ \bvY \mid I_{i-1} \leq \Re(\bvY) < I_{i} \,, Q_{j-1} \leq \Im(\bvY) < Q_{j} \}\,,
\end{multline}
for $i,j\in\{1,\ldots, S\}$, where $I_1,\ldots,I_{S-1}$ and $Q_1,\ldots,Q_{S-1}$ are the thresholds for in-phase and quadrature, respectively. We set $I_0 = -\infty$, $I_S = \infty$, $Q_0 = -\infty$, and $Q_S = \infty$ in order to go along with the unified notation.
  \item \textbf{Amplitude/phase quantization:} As for $2\times2$ LoS MIMO, the quantization set for this scheme is specified as in \eqref{eq:Amp_Quant_Region}.
\end{itemize}

We determine the quantizer regions (i.e., the thresholds in \eqref{eq:Amp_Quant_Region} and \eqref{eq:I_Q_Region}) based on the following two different metrics:
\begin{itemize}
  \item \textbf{Minimum mean squared quantization error (MMSQE)-based regions:} This is the conventional approach to quantizer design based on minimization of the mean squared error given by
  \begin{align}\label{eq:MMSQE_design}
  \mathop{\mathbb{E}}\{(\tilde{Y} - Q(\tilde{Y}))^2\}
  \end{align}
  where $Q(\cdot)$ is the quantizer function, whose set is defined as either \eqref{eq:Amp_Quant_Region} or \eqref{eq:I_Q_Region}. The optimal decision boundaries in \eqref{eq:Amp_Quant_Region} and \eqref{eq:I_Q_Region} are obtained as usual, by applying the Lloyd-Max algorithm \cite{Lloyd_82,Max_60}.

  \item \textbf{Equal probability-based regions:} The quantizer regions here are obtained by partitioning the fitted complex Gaussian distribution into equal probability regions.
  In other words, the quantizer boundaries maximize the entropy of $\tilde{Y}$; that is, $H(\tilde{Y})$. For the circular Gaussian distribution, the boundaries can
  be specified analytically.
  For I/Q quantization, the entropy-maximizer thresholds can be calculated as
  \begin{align}\label{eq:IQ_Thresholds}
	 I_i = Q_i = \tilde{\mu} + \frac{\tilde{\sigma}}{\sqrt{2}}\inv{\Phi}\left(\frac{i}{S}\right) = \sqrt{\frac{1 + \sigma^2}{2}}\inv{\Phi}\left(\frac{i}{S}\right)
  \end{align}
  for $i \in\{1,\ldots, S-1\}$, where $\inv{\Phi}$ is the inverse distribution function (i.e., the quantile function) for the standard Gaussian distribution with a mean of $0$ and a standard deviation of $1$.
  For amplitude/phase quantization, the phase quantization is uniform, and the amplitude thresholds that maximize the entropy can be found as
  \begin{align}\label{eq:AP_Thresholds}
	 A_i = \sqrt{(1 + \sigma^2)\log{\left(\frac{K}{K-i}\right)}}
  \end{align}
  for $i \in\{1,\ldots, K-1\}$.
\end{itemize}

For those two different metrics, Fig.~\ref{fig:Quantization_4x4} shows the I/Q and amplitude/phase quantizers at 10 dB SNR, each having a total of $16$ regions (i.e., $K=2$, $M = 8$, and $S = 4$). 

\subsection{Numerical Results}\label{sec:NumRes_4by4_Quantization}

\begin{figure}[htbp]
\vspace{-0.2cm}
\begin{center}
\begin{subfigure}[b]{0.48\textwidth}
\includegraphics[width=\columnwidth,draft=false]{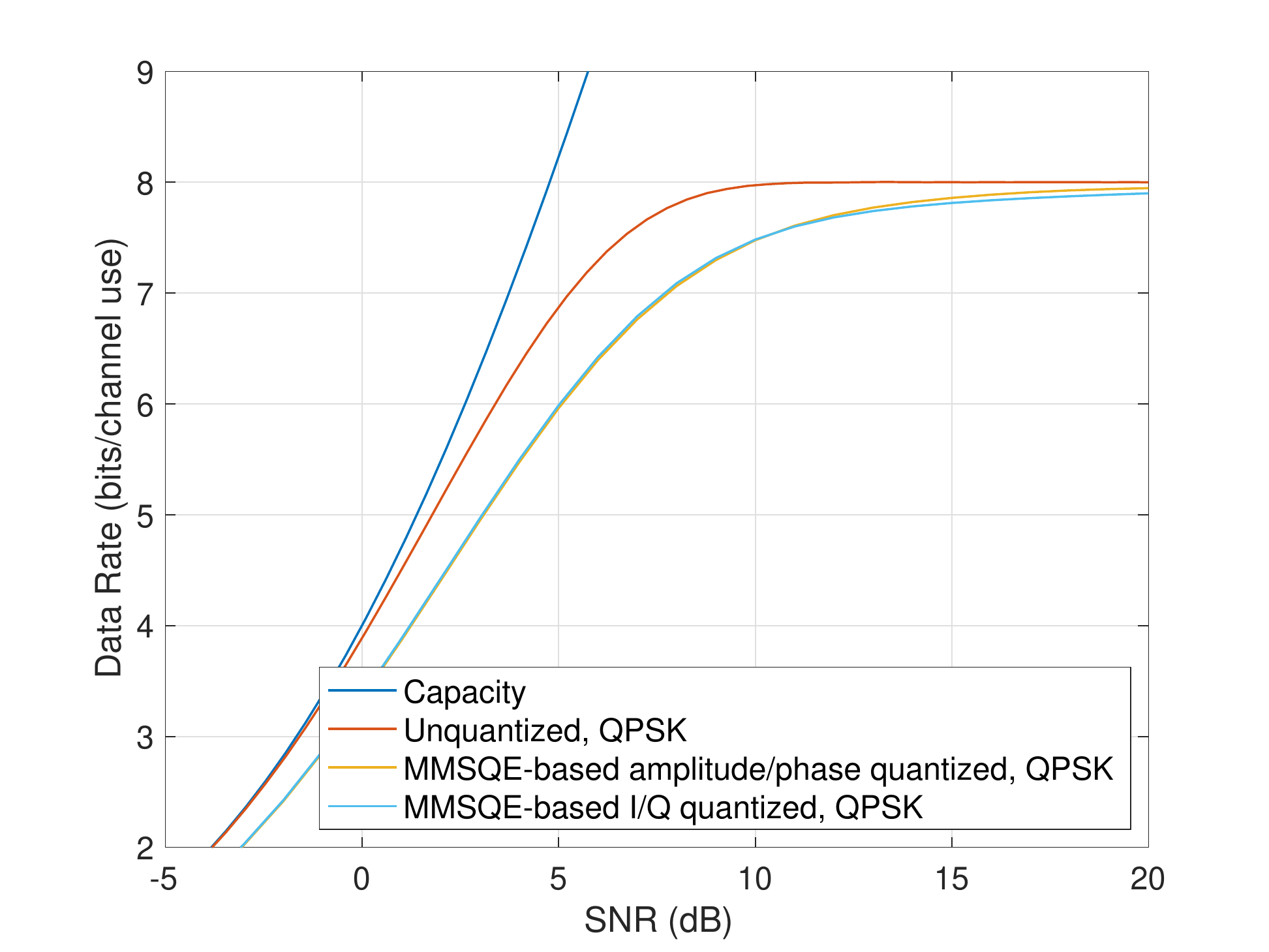}
\caption{}\label{fig:IQ_and_AP_MMSQE}
\end{subfigure}
\begin{subfigure}[b]{0.48\textwidth}
\includegraphics[width=\columnwidth,draft=false]{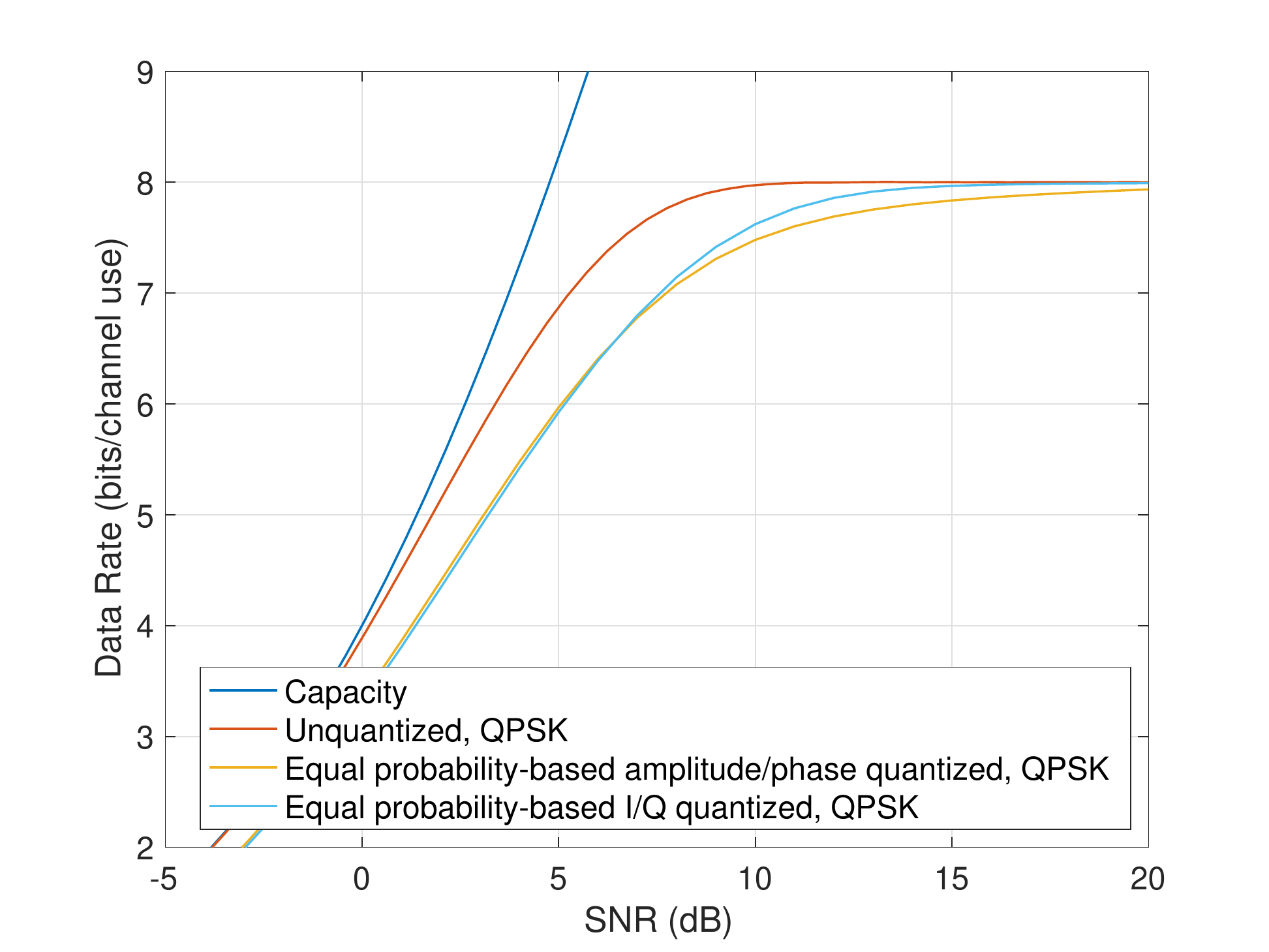}
\caption{}\label{fig:IQ_and_AP_EqualProb}
\end{subfigure}
\caption{Data rate versus SNR for (a) MMSQE-based quantization and (b) equal probability-based quantization for a $4\times4$ system.}
\end{center}
\vspace{-0.6cm}
\end{figure}

We now investigate Shannon limits for different quantizer designs; in each case, there are 16 quantization regions in the complex plane.
Fig.~\ref{fig:IQ_and_AP_MMSQE} shows that, for conventional MMSQE-based quantization, amplitude-phase and I/Q quantization are both unable to
reach the maximum data rate ($8$ bits per channel use) even at SNR as high as 20 dB, compared to the unquantized system, which saturates at 10 dB SNR. 
On the other hand, Fig.~\ref{fig:IQ_and_AP_EqualProb} shows that, for equal probability regions, I/Q quantization attains the maximum data rate of $8$ bits per channel use 
at around 15 dB, while amplitude/phase quantization continues to exhibit a gap to the unquantized limit even at SNR of 20 dB.  We conclude that 2 bit quantization on I and Q based on equal probability regions should suffice to attain acceptable performance at moderate SNR, and focus on
this setting for investigation of spatial demultiplexing algorithms in the next section.

%% file: detection_virtual_quantization.tex
\section{Spatial Demultiplexing under Severe Quantization}\label{sec:SpatialDemultiplexing}

We begin with spatial demultiplexing for $4 \times 4$ LoS MIMO system with QPSK modulation with 2 bit I/Q quantization as designed in the previous section.  
In order to highlight the impact of quantization, consider an ideally conditioned channel with $\theta = \frac{\pi}{2}$ in (\ref{eq:ChannelMatrix_4x4}), for which 
the received antenna responses for different transmitted streams are orthogonal, so that matched filter, linear ZF and maximum likelihood detection all yield the same
performance with {\it unquantized} observations.  We shall see, however, that drastic quantization can have a severe impact on the performance of linear detection even in such an ideal setting because of the common channel phase $\Phi$ in (\ref{eq:ChannelMatrix_4x4}), which can move the observations close to quantization boundaries.  On the other hand, the mutual information plot in
Figure \ref{fig:IQ_and_AP_EqualProb} shows that reliable communication at the maximum data rate of 8 bits per channel use should be possible at an SNR of about 15 dB in this setting.
Our goal, therefore, is to devise spatial demultiplexing with reasonable complexity that can approach this performance. 
We consider an uncoded BER target of about $10^{-3}$, which yields reliable communication with high-rate bit interleaved
coded modulation, since a binary symmetric channel with this cross-over probability has capacity close to 0.99 bits per channel use.

From the point of view of minimizing the probability of error, the optimal detector based on the quantized output is the maximum likelihood (ML) detector \cite{Yang_Survey_2015}, given by
  \begin{align}\label{eq:ML_detector}
	 \vXH(\vYQ) = \mathop{\underset{\vX \in \mathcal{S}^{N_R}}{\mathrm{argmax}}}\,p(\vYQ \mid \vX)\,.
  \end{align}
Since the minimization in \eqref{eq:ML_detector} is over all possible transmitted vectors and it is difficult to calculate $p(\vYQ \mid \vX)$ for a given $\vX$, the problem in \eqref{eq:ML_detector} has prohibitive complexity. As a low-complexity alternative, we consider linear ZF detection.  
We have verified by simulations that other standard multiuser detection techniques such as linear minimum mean squared error (MMSE) detector \cite{Honig_Book_2009} and sphere decoding \cite{Damen_2003} achieve the same performance as linear ZF for our system.

Recall that the ZF solution minimizes 
  \begin{align}\label{eq:ZF_problem}
	 \tilde{\vX}(\vYQ) = \underset{\vX}{\mathrm{argmin}} \norm{\vYQ - \vH\vX}
  \end{align}
 and can be obtained by computing 
  \begin{align}\label{eq:ZF_step1}
	 \tilde{\vX}(\vYQ) = \inv{(\vH^\dagger \vH)}\vH^\dagger\vYQ
  \end{align}
first, and then finding $i$th element of $\vXH(\vYQ)$ as
  \begin{align}\label{eq:ZF_step2}
	 \hat{X}_i(\vYQ) = \underset{X_i \in \mathcal{S}}{\mathrm{argmin}} |\tilde{X}_i(\vYQ) - X_i|
  \end{align}
for all $i\in\{1,\ldots, 4\}$, where $\tilde{X}_i(\cdot)$ is the $i$th element of $\tilde{\vX}(\cdot)$. Note that $\vYQ$ in \eqref{eq:ZF_step1} represents the quantized output at the receiver and the quantizer outputs are set to the centroids of the quantizer regions, which are obtained based on the complex Gaussian approximation. Mathematically, the quantized output of $i$th antenna for the received signal $Y_i$ is equal to 
  \begin{align}\label{eq:Centroids}
  	\bar{Y}_i = \frac{\int_{\tilde{\Gamma}_i} y f_{\tilde{Y}}(y)\,dy}{\int_{\tilde{\Gamma}_i} f_{\tilde{Y}}(y)\,dy}\,,
  \end{align}
where $\tilde{\Gamma}_i$ denotes the quantizer region that $Y_i$ falls into; that is, $Y_i \in \tilde{\Gamma}_i$.

\subsection{Virtual Quantization}\label{sec:VirQuantization}

Linear ZF detection based on the centroids codebook performs poorly when the unquantized outputs are far from the centroids of the regions that they belong to.  On the other hand,
we know that linear ZF detection with unquantized outputs yields excellent performance for a well-conditioned MIMO channel.  This motivates viewing
the unquantized output vector as a hidden variable, or nuisance parameter, for our hypothesis testing problem of estimating the transmitted symbols.
We can now apply any of the standard tools of composite hypothesis testing to this problem.  We choose here a Generalized Likelihood Ratio Test (GLRT) approach,
in which we jointly estimate the unquantized output and the transmitted symbols given the quantized outputs. 
 That is, the transmitted symbols are estimated by jointly maximizing the probability of unquantized output and transmitted symbols given the quantized output:
  \begin{align}\label{eq:Joint_Estimation}
	 (\vXH(\vYQ), \vY(\vYQ)) = \mathop{\underset{\vX \in \mathcal{S}^{N_R}, \vY}{\mathrm{argmax}}}\,p(\vX, \vY \mid \vYQ)\,.
  \end{align}
A key difficulty in this optimization problem is that we need to search over a continuum of values for the hidden unquantized output $\vY$. 
We therefore consider a grid-based approximation of $\vY$, where the grid is finer than that provided by the quantizer.  We term this approach
{\it virtual quantization.}

\begin{figure}[htbp]
\vspace{-0.2cm}
\begin{center}
\includegraphics[width=0.7\columnwidth,draft=false]{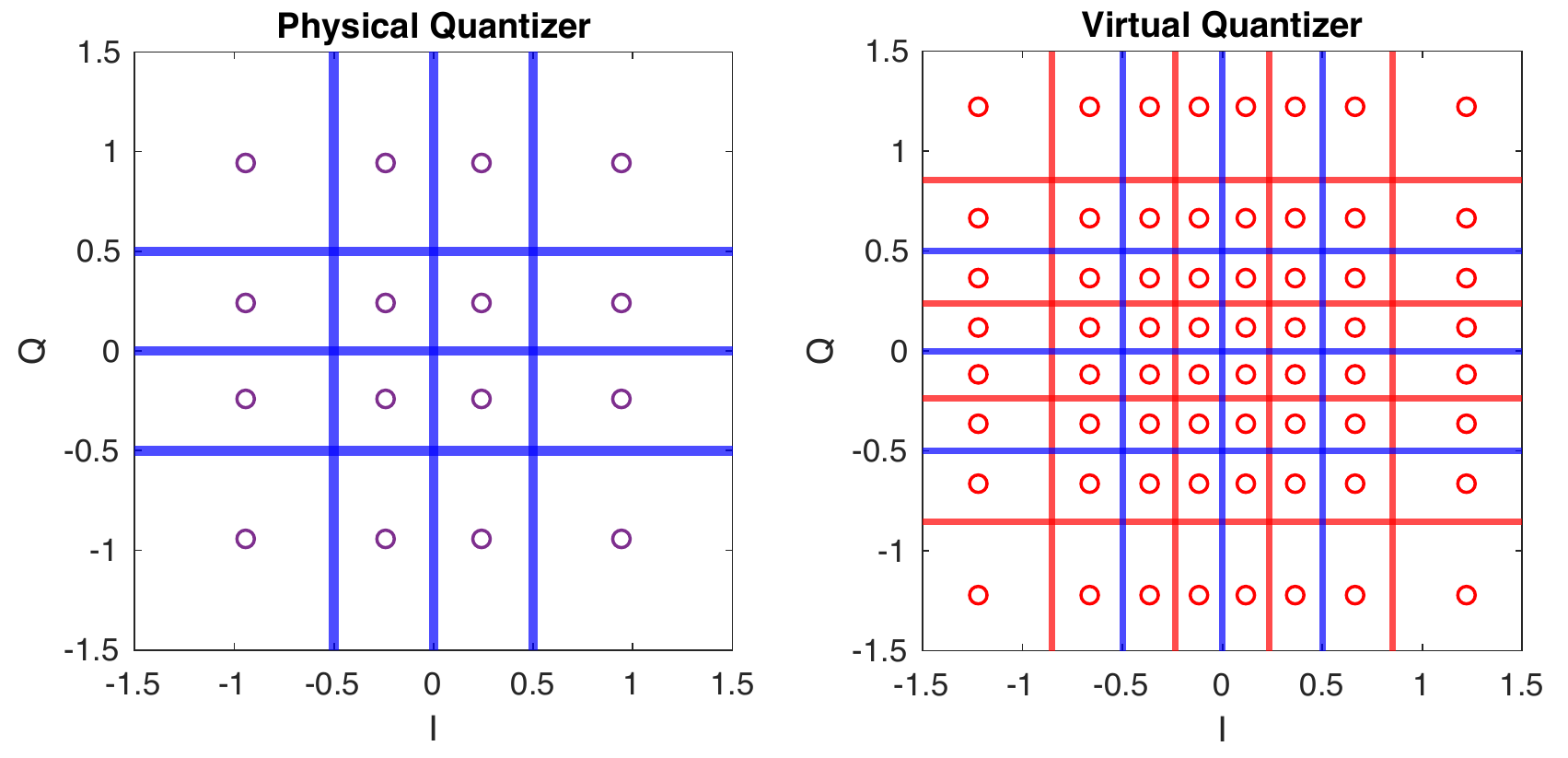}
\vspace{-0.3cm}
\caption{Physical and virtual quantizers with the outputs corresponding to the centroids of the regions for $10\,$dB SNR.}\label{fig:Virtual_quantizer}
\end{center}
\vspace{-0.6cm}
\end{figure}

\begin{figure*}
\vspace{-0.2cm}
\begin{center}
\includegraphics[width=0.7\columnwidth,draft=false]{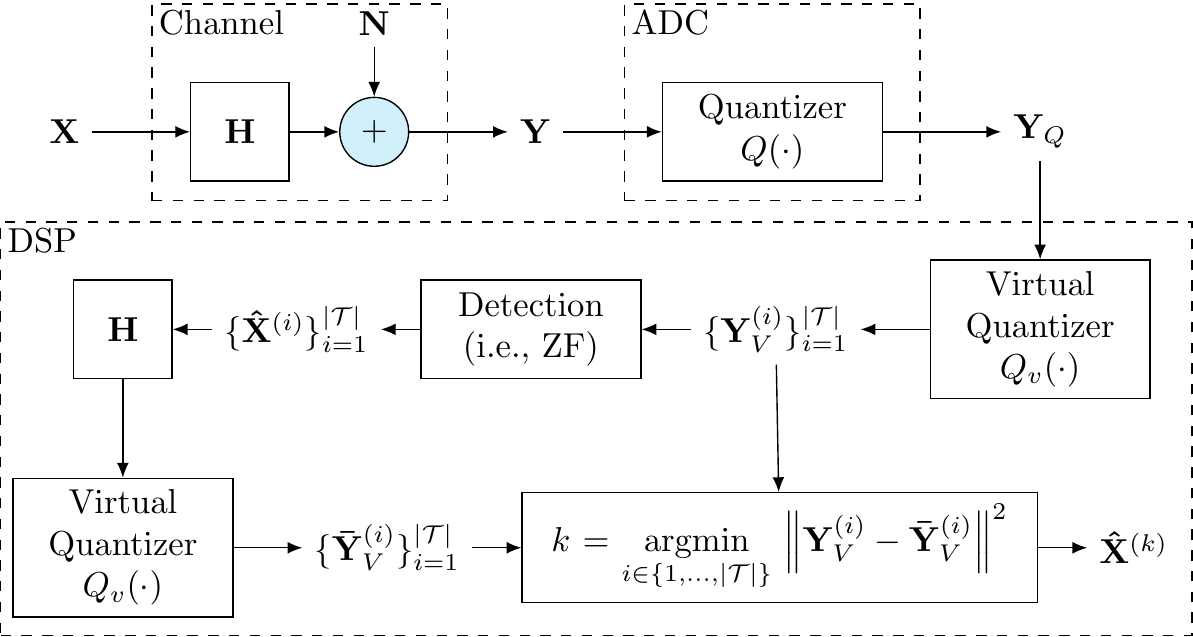}
\vspace{-0.3cm}
\caption{Flow diagram of virtual quantization based spatial demultiplexing.}\label{fig:Flow_Virtual_quantizer}
\end{center}
\vspace{-0.6cm}
\end{figure*}

Let $Q_v(\cdot)$ denote the element-wise virtual quantizer function. The virtual quantized hidden variable can be obtained as $\vYV = Q_v(\vY)$. Then, 
the quantized output $\vYQ$ can be written as $\vYQ = Q_c(\vYV)$ as a coarsening of virtual quantized hidden variable, where $Q_c(\cdot)$ is the coarsening function. Thus, $\vYQ = Q_c(Q_v(\vY)) = Q(\vY)$, where $Q(\cdot)$ is the element-wise actual quantizer function defined in \eqref{eq:Quantizer} for the $i$th receive antenna and $Q_c(\cdot)$ can be considered as $Q_c(\cdot) = Q(\cdot)$. 
Thus, $\vX \rightarrow \vY \rightarrow \vYV \rightarrow \vYQ$ is a Markov chain. Figure \ref{fig:Virtual_quantizer} shows an example of physical and virtual quantizers used in our numerical results (detailed description is provided later in this section).

Quantizing the nuisance parameter $\vY$ in the joint estimation problem in \eqref{eq:Joint_Estimation} using the virtual quantizer, we seek to jointly estimate
the transmitted symbols and the virtual quantizer outputs:
  \begin{align}\label{eq:Joint_Estimation_Approx}
	 (\vXH(\vYQ), \vY(\vYQ)) &= \mathop{\underset{\vX \in \mathcal{S}^{N_R}, \vY}{\mathrm{argmax}}}\,p(\vX, Q_v(\vY) \mid \vYQ) \\\label{eq:Joint_Estimation_Approx_2}
	 & = \mathop{\underset{\vX \in \mathcal{S}^{N_R}, \vYV \in \mathcal{T}}{\mathrm{argmax}}}\,p(\vX, \vYV \mid \vYQ)
  \end{align}
where \eqref{eq:Joint_Estimation_Approx_2} follows from $\vYV = Q_v(Y)$ and $\mathcal{T} = \mathcal{T} \left( \vYQ \right)$ is a discrete set including all possible combinations of virtual quantized output for observed $\vYQ$:
$$
\mathcal{T} = \{\vYV = Q_v(\vY) \mid Q(\vYV) = \vYQ \}
$$

In order to solve the problem in \eqref{eq:Joint_Estimation_Approx_2} and estimate the transmitted symbols, we maximize the objective function in \eqref{eq:Joint_Estimation_Approx_2} with respect to $\vX$ for a given $\vYV \in \mathcal{T}$ first, then substitute the obtained $\vX$ to the objective function to solve for $\vYV$. To begin with, for a given $\vYV \in \mathcal{T}$, the maximization of the optimization problem in \eqref{eq:Joint_Estimation_Approx_2} over $\vX$ can be expressed as follows:
\begin{align}\label{eq:Joint_prob_max_over_X_1}
    \vXH(\vYV, \vYQ) &= \mathop{\underset{\vX \in \mathcal{S}^{N_R}}{\mathrm{argmax}}}\,p(\vX, \vYV \mid \vYQ) \\\label{eq:Joint_prob_max_over_X_2}
    & = \mathop{\underset{\vX \in \mathcal{S}^{N_R}}{\mathrm{argmax}}}\,p(\vX \mid \vYV, \vYQ)\,p(\vYV \mid \vYQ) \\\label{eq:Joint_prob_max_over_X_3}
    & = \mathop{\underset{\vX \in \mathcal{S}^{N_R}}{\mathrm{argmax}}}\,p(\vX \mid \vYV)\,p(\vYV \mid \vYQ) \\\label{eq:Joint_prob_max_over_X_4}
    & = \mathop{\underset{\vX \in \mathcal{S}^{N_R}}{\mathrm{argmax}}}\,p(\vX \mid \vYV) \\\label{eq:Joint_prob_max_over_X_5}
    & = \mathop{\underset{\vX \in \mathcal{S}^{N_R}}{\mathrm{argmax}}}\,p(\vYV \mid \vX)
  \end{align}
where \eqref{eq:Joint_prob_max_over_X_2}  is a standard conditional probability computation, \eqref{eq:Joint_prob_max_over_X_3} 
is based on the Markov property $\vX \rightarrow \vYV \rightarrow \vYQ$, \eqref{eq:Joint_prob_max_over_X_4} is obtained since $p(\vYV \mid \vYQ)$ does not depend on $\vX$ for given $\vYV$ and $\vYQ$, and \eqref{eq:Joint_prob_max_over_X_5} follows from Bayes' theorem and equally likely transmitted symbols.

The optimization problem in \eqref{eq:Joint_prob_max_over_X_5} is in the form of \eqref{eq:ML_detector}, except that it considers virtual quantized hidden variable instead of actual quantized output. As described in \eqref{eq:ZF_step1} and \eqref{eq:ZF_step2}, \eqref{eq:Joint_prob_max_over_X_5} can be approximated by linear ZF detection. 
Therefore, for given $\vYV$ and observed quantized output $\vYQ$, $\vXH(\vYV, \vYQ)$ can be obtained as in \eqref{eq:ZF_step1} and \eqref{eq:ZF_step2}.

Next, we substitute $\vXH(\vYV, \vYQ)$ to the optimization problem in \eqref{eq:Joint_Estimation_Approx_2} and maximize it over $\vYV$. Substituting $\vXH(\vYV, \vYQ)$ to \eqref{eq:Joint_Estimation_Approx_2},
  \begin{align}\label{eq:Joint_prob_max_over_Y_v_1}
    \hat{Y}_v(\vXH, \vYQ) &= \mathop{\underset{\vYV \in \mathcal{T}}{\mathrm{argmax}}} \,p(\vXH, \vYV \mid \vYQ) \\\label{eq:Joint_prob_max_over_Y_v_2}
    & = \mathop{\underset{\vYV \in \mathcal{T}}{\mathrm{argmax}}}\,p(\vYV \mid \vXH, \vYQ)\,p(\vXH \mid \vYQ) \\\label{eq:Joint_prob_max_over_Y_v_3}
    & \approx \mathop{\underset{\vYV \in \mathcal{T}}{\mathrm{argmin}}}\,\norm{\vYV - \vYVH}^2 
  \end{align}
where \eqref{eq:Joint_prob_max_over_Y_v_2} follows from Bayes' theorem and \eqref{eq:Joint_prob_max_over_Y_v_3} is obtained by approximating the impact of quantization and noise as Gaussian and modeling $p(\vYV \mid \vXH, \vYQ)$ as Gaussian with mean $\vYVH$, where $\vYVH$ denotes the noiseless reconstruction and is equal to $\vYVH = Q_v(\vH\vXH)$. 
Note that we ignore $p(\vXH \mid \vYQ)$ term in \eqref{eq:Joint_prob_max_over_Y_v_3} even though $\vXH$ depends on $\vYV$, which means that we are not necessarily
attaining the true maximum for (\ref{eq:Joint_prob_max_over_Y_v_1}). We expect the impact of this approximation on  \eqref{eq:Joint_prob_max_over_Y_v_3} to be small 
compared to the term $p(\vYV \mid \vXH, \vYQ)$, which decays exponentially according to our Gaussian approximation for the sum of the virtual quantization noise and thermal noise.

The proposed virtual quantization method can now be summarized as follows:
\begin{enumerate}
  \item Generate set $\mathcal{T}$ based on the physical quantized observation and the virtual quantizer function such that $\mathcal{T} = \{\vYV = Q_v(\vY) \mid Q(\vYV) = \vYQ \}$.
  \item For each $\vYV \in \mathcal{T}$, calculate the ZF solution to obtain $\vXH(\vYV, \vYQ)$ by treating $\vYV$ as if it is the observed output at the receiver.
  \item Find $\vYV \in \mathcal{T}$ that minimizes the Euclidean distance between $\vYV$ and $Q_v(\vH\vXH)$; that is, $\lVert \vYV -Q_v(\vH\vXH)\rVert^2$.
  \item Declare the corresponding $\vXH(\vYV, \vYQ)$ as the estimated symbol vector. 
\end{enumerate}
Fig.~\ref{fig:Flow_Virtual_quantizer} illustrates this procedure via a flow diagram. 

The result of the estimation depends on the virtual quantization function (i.e., $Q_v(\cdot)$) and the virtual quantization output (i.e., $\vYV$), both of which determine the elements in $\mathcal{T}$ and contribute to the cost function used to specify the estimated symbols. Different virtual quantization functions can be designed for our proposed method. However, in this paper, we consider the same method that we use for the design of the actual physical quantizer. Considering the Gaussian approximation discussed in the design of the actual quantizer, we design an I/Q quantization-based virtual quantizer having equal-probability regions. For the 2-bit per I/Q physical quantizer having $16$ regions, we consider a virtual quantizer with $64$ regions (i.e., $S = 8$) whose outputs are decided based on the centroids of the corresponding regions similar to those of the physical quantizer, as shown in Fig.~\ref{fig:Virtual_quantizer}. In this case, the virtual quantizer divides each physical quantizer bin into $4$ regions, which can be considered as a virtually created 1-bit quantizer per each I/Q for each physical quantization bin. 

\subsection{Numerical Results}\label{sec:NumRes_Demodulation}

\begin{figure}[htbp]
\vspace{-0.2cm}
\begin{center}
\begin{subfigure}[b]{0.48\textwidth}
\includegraphics[width=\columnwidth,draft=false]{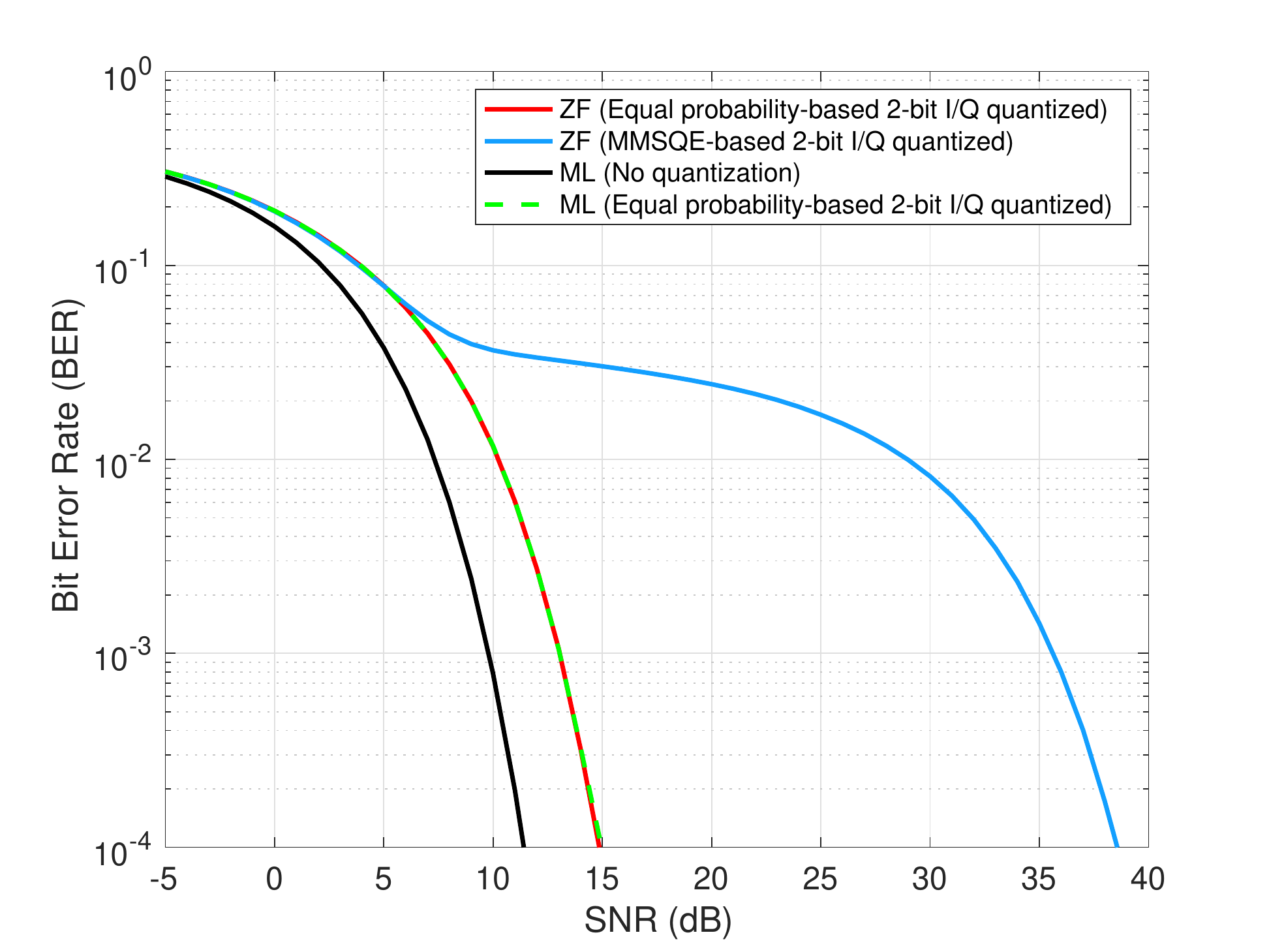}
\caption{}\label{fig:Detection_phi_zero}
\end{subfigure}
\begin{subfigure}[b]{0.48\textwidth}
\includegraphics[width=\columnwidth,draft=false]{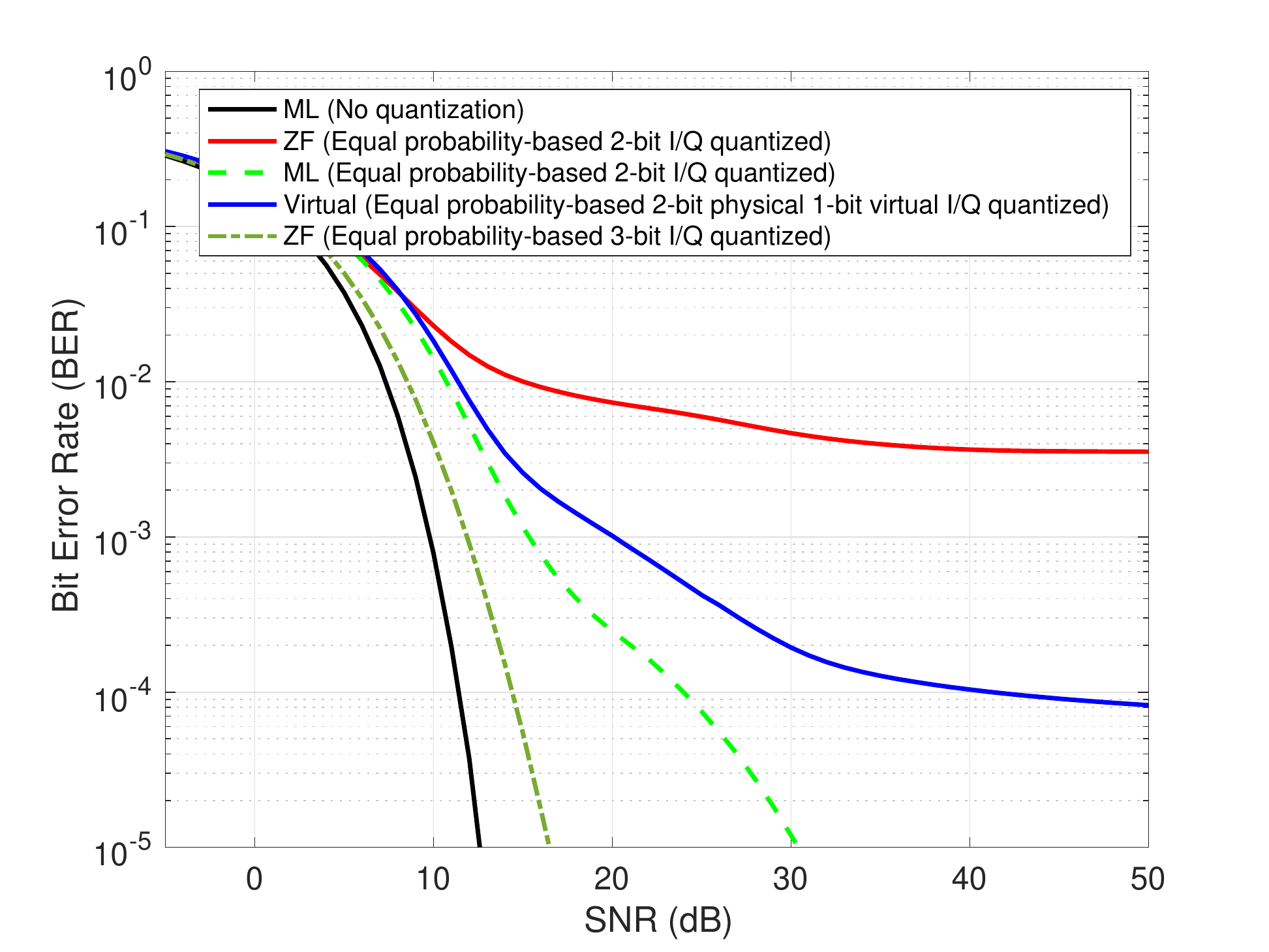}
\caption{}\label{fig:Detection_phi_average}
\end{subfigure}
\caption{Bit error rate versus SNR for different detection methods when the common phase is (a) fixed to $\Phi = 0$; (b) uniformly distributed over $[0, 2\pi)$, for a $4\times4$ MIMO system with QPSK.}
\end{center}
\vspace{-0.6cm}
\end{figure}

We consider two benchmarks: ML with no quantization and ML with quantization. The former assumes there is no quantization in the system; that is $\vYQ = \vY$ and provides an unquantized benchmark as if ADCs at the receiver have infinite precision. The latter considers the quantized outputs at the receiver and is obtained based on \eqref{eq:ML_detector}.

For a $4\times4$ MIMO system, Fig.~\ref{fig:Detection_phi_zero} plots bit error rate (BER) versus SNR, setting $\theta = \pi/2$, for different detection methods when the common phase is fixed to $\Phi = 0$. 
We may also view this as equivalent to a hybrid analog-digital processing scheme in which the common channel phase $\Phi$ is removed by analog derotation {\it prior} to quantization.
The plot indicates that linear ZF detection with equal probability quantization and the centroids codebook remains near-optimal, achieving the same performance as ML reception,
 as long as the common channel phase is removed prior to quantization. On the other hand, as expected from our mutual information computations, MMSQE-based quantization performs
 significantly worse than the equal probability-based quantizer.  Note that we have simulated other detection methods such as the linear MMSE detector and the sphere decoder based on the quantized outputs at the receiver, and have verified that they achieve the same performance as linear ZF detection.

We now turn to the scenario of interest for us: fully digital processing without derotation of the common phase $\Phi$ prior to quantization. The receiver processing with quantized observations does employ
knowledge of $\Phi$, but performance is adversely affected by points being rotated close to quantization boundaries. Fig.~\ref{fig:Detection_phi_average} plots BER averaged over the common phase $\Phi$ versus SNR, setting $\theta = \pi/2$, for different detection methods, where $\Phi$ is uniformly distributed over $[0, 2\pi)$. We now see that linear ZF detection with the centroids codebook performs significantly worse, with an error floor that stays higher than our target $10^{-3}$ BER. The proposed virtual quantization approach performs significantly better: its performance is close to that of ML detection at BER of $10^{-2}$, while being 5 dB worse at the target BER of $10^{-3}$.  It still exhibits an error floor at $10^{-4}$, motivating additional effort in devising low-complexity strategies for approaching maximum likelihood performance. 

\begin{figure}[htbp]
\vspace{-0.2cm}
\begin{center}
\includegraphics[width=0.7\columnwidth,draft=false]{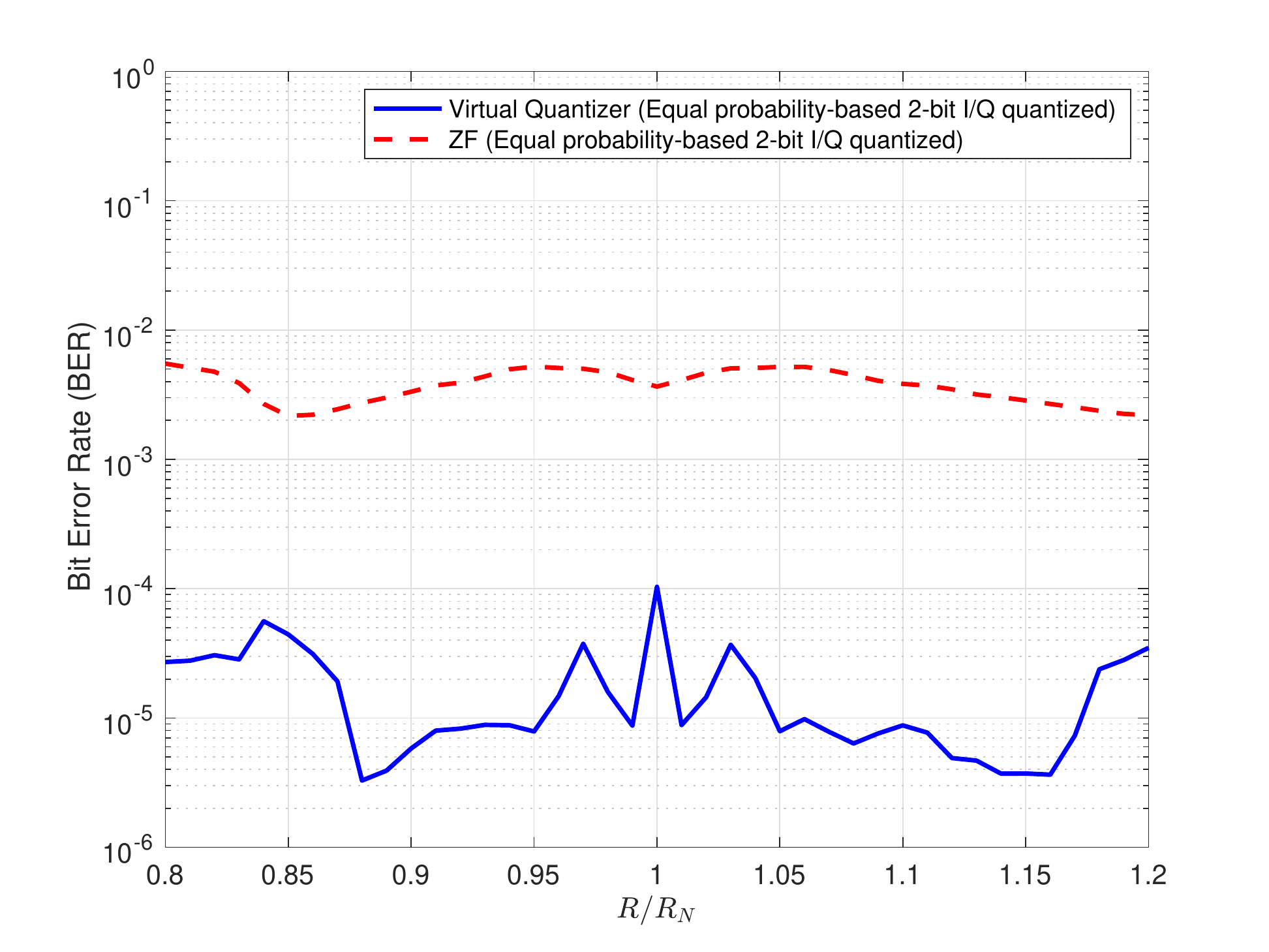}
\vspace{-0.3cm}
\caption{Bit error rate averaged over $\Phi$ versus $R/R_N$ for different detection methods for a $4\times4$ MIMO system with QPSK, where SNR is $40\,$dB.}\label{fig:BER_vs_RangeChanges}
\end{center}
\vspace{-0.6cm}
\end{figure}

\noindent
{\bf Beyond the ideal model:} Since severe quantization destroys the orthogonality of the received signals for different data streams, we expect that our all-digital receiver should
be robust to changes in link distance $R$ around the nominal range $R_N$.  Fig.~\ref{fig:BER_vs_RangeChanges} plots BER vs $R/R_N$ at $40\,$dB SNR for our proposed virtual quantization approach and linear ZF detection with the centroids codebook, where $R \in [0.8R_N, 1.2R_N]$. The poor performance of linear ZF detection also persists as we vary $R$.

\begin{figure}[htbp]
\vspace{-0.2cm}
\begin{center}
\includegraphics[width=0.7\columnwidth,draft=false]{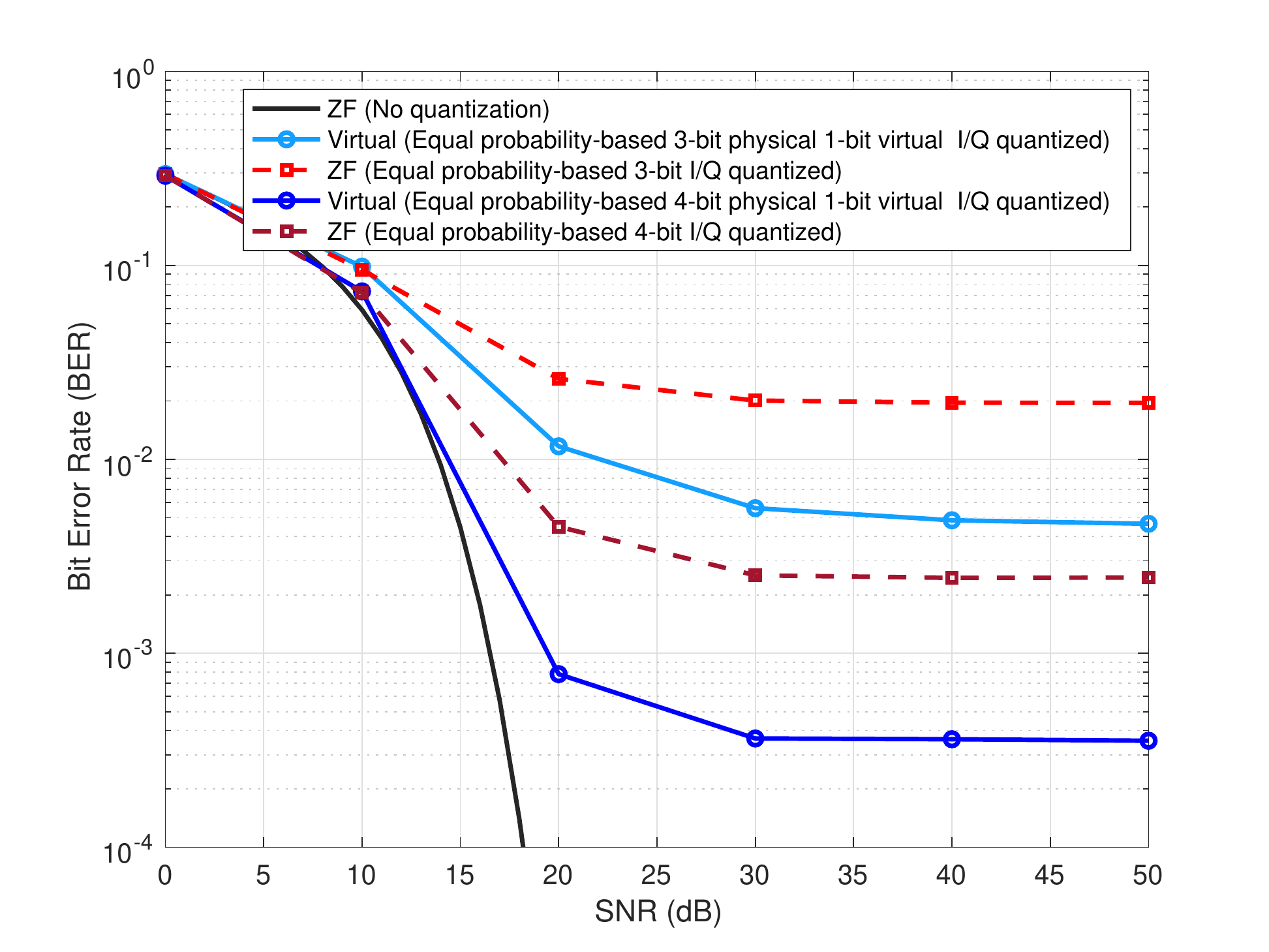}
\vspace{-0.3cm}
\caption{Bit error rate averaged over $\Phi$ versus SNR for different detection methods including our proposed virtual quantization approach when the common phase is uniformly distributed over $[0, 2\pi)$ for a $4\times4$ MIMO system with 16QAM.}\label{fig:Detection_phi_average_16QAM}
\end{center}
\vspace{-0.6cm}
\end{figure}

\noindent
{\bf Scaling to larger constellations:} A key advantage of virtual quantization is that its complexity does not scale with constellation size.
We verify this by evaluating performance for 16QAM modulation.  Given the higher dynamic range of 16QAM, we consider 3 bit and 4 bit I/Q physical quantization, and then
add 1 bit virtual quantization as before. Fig.~\ref{fig:Detection_phi_average_16QAM} plots BER averaged over the common phase $\Phi$ versus SNR, setting $\theta = \pi/2$, for different detection methods. We see that neither 4 bit physical quantization (without virtual quantization) nor 3 bit physical quantization with 1 bit virtual quantization achieve our
target $10^{-3}$ BER even at very high SNR.  However, 4 bit physical quantization with 1 bit virtual quantization does achieve our BER target at approximately $20\,$dB SNR.

%% file: conclusion.tex
\section{Conclusion} \label{sec:conclusion}

Our study of ideal $4 \times 4$ LoS MIMO system at high SNR yields fundamental insight into the impact of severe quantization.  We show that equal probability quantization, which maximizes per-antenna output entropy, outperforms
standard MMSQE quantization in such regimes.  For spatial demultiplexing, we introduce the novel concept of virtual quantization, which may be viewed as approximate joint estimation of the transmitted symbols and the unquantized received signal, and show that linear detection with virtual quantization is an effective low-complexity alternative to maximum likelihood detection, which requires complexity exponential in the number of transmitted bits. It remains an open issue as to whether the gap to maximum likelihood detection at higher SNR can be further reduced, and the error floor eliminated, while maintaining reasonable complexity. 

An important direction for future research is to investigate quantization-constrained LoS MIMO is more complex settings, including understanding the impact of dispersion due to geometric misalignments and potential performance advantages
of spatial oversampling \cite{Sawaby_Asilomar2016, Raviteja_Spawc2018}. While we do not consider transmit precoding here, joint transmit-receive optimization subject to dynamic range constraints and nonlinearities at both ends is of great interest. 
At a fundamental level, the concept of virtual quantization, which treats the unquantized output as a hidden variable, may be worth exploring for other system models.